\documentclass[preprint,showpacs,preprintnumbers,amsmath,amssymb,nofootinbib]{revtex4}

\usepackage{etex}

\usepackage{amsthm,amscd,amsbsy,array}
\usepackage{bm}
\usepackage{soul} 

\usepackage{graphics,graphicx,xcolor}

\usepackage[colorlinks=true, pdfstartview=FitV, linkcolor=blue, citecolor=blue, urlcolor=blue]{hyperref} 

\newcommand{\gb}{\colorbox{green}}

\newenvironment{redtext}{\color{red}}
{\ignorespacesafterend}
\newenvironment{bluetext}{\color{blue}}{\ignorespacesafterend}

\newenvironment{magentatext}{\color{magenta}}{\ignorespacesafterend}
\newenvironment{cyantext}{\color{cyan}}{\ignorespacesafterend}

\newcommand{\bmagenta}{\begin{magentatext}}
\newcommand{\emagenta}{\end{magentatext}}
\newcommand{\bcyan}{\begin{cyantext}}
\newcommand{\ecyan}{\end{cyantext}}
\newcommand{\bblue}{\begin{bluetext}}
\newcommand{\eblue}{\end{bluetext}}
\newcommand{\bred}{\begin{redtext}}
\newcommand{\ered}{\end{redtext}}

\numberwithin{equation}{section}

\let\ssection=\section
\renewcommand{\section}{\setcounter{equation}{0}\ssection}
\newcommand{\ben}{\begin{equation}}
\newcommand{\een}{\end{equation}}

\newcommand{\cA}{{\mathcal{A}}}

\newcommand{\bb}{{\bf b}}

\newcommand{\bc}{{\mathbf{c}}}

\newcommand{\cC}{{\mathcal{C}}}

\newcommand{\Carr}{{\mathrm{Carr}}}

\def\aand{{\quad\text{and}\quad}}
\def\where{{\quad\text{where}\quad}}

\newcommand{\cD}{{\mathcal{D}}}

\newcommand{\diag}{\mathrm{diag}}

\newcommand{\cF}{{\mathcal{F}}}

\newcommand{\cH}{{\mathcal{H}}}

\newcommand{\Id}{\mathrm{Id}}

\newcommand{\cL}{{\mathcal{L}}}

\newcommand{\cM}{\mathcal{M}}

\newcommand{\cN}{\mathcal{N}}

\newcommand{\bp}{{\bf p}}

\newcommand{\bu}{\mathbf{u}}

\newcommand{\bx}{{\bm{x}}}

\newcommand{\SO}{\mathrm{SO}}
\newcommand{\ISO}{\mathrm{ISO}}
\newcommand{\SIM}{\mathrm{SIM}}
\newcommand{\ISIM}{\mathrm{ISIM}}
\newcommand{\DISIM}{\mathrm{DISIM}}

\newcommand{\bv}{{\bf v}}

\def\smallover#1/#2{\hbox{$\textstyle\frac{#1}{#2}$}} %

\def\bu{{\bm{u}}}
\def\bv{{\bm{v}}}
\def\bp{{\bm{p}}}

\def\parag{\hfil\break} 
\def\kikezd{\parag\underbar}

\def\benu{\begin{enumerate}}
\def\eenu{\end{enumerate}}
\def\beq{\begin{equation}}
\def\eeq{\end{equation}}
\def\beqa{\begin{eqnarray}}
\def\eeqa{\end{eqnarray}}
\def\nn{\nonumber}
\def\barray{\left(\begin{array}}
\def\earray{\end{array}\right)}
\def\barraynb{\begin{array}}
\def\earraynb{\end{array}}

\def\?{\quad{\gb{\fbox{\texttt{?}}\;}}\quad}
\def\p{{\partial}}

\def\v0{\mathbf{0}}

\def\Rarrow{{\quad\Rightarrow\quad}}
\def\B-F{{Bogoslovsky-Finsler}}
\def\FL{{Friedmann-Lema\^{\i}tre\,}}

\def\beq{\begin{equation}}
\def\eeq{\end{equation}}
\def\bea{\begin{eqnarray}}
\def\eea{\end{eqnarray}}

\def\p{\partial}

\def \p{{\partial}}

\def\6{\partial}
\def\7{\tilde}
\def\8{\widehat}
 \def\bx{{\bf x}}

\def\G11{\Gamma_{11} }

\newcommand{\const}{\mathop{\rm const.}\nolimits}
\newcommand{\half }{\smallover{1}/{2}}

\def\smallover#1/#2{\hbox{$\textstyle\frac{#1}{#2}$}} %
\def\smallcirc{{\raise 0.5pt \hbox{$\scriptstyle\circ$}}}
\def\2{{\smallover1/2}}

\newcommand{\bigbox}[1]{\fbox{%
\rule[-20pt]{0pt}{45pt}$\;\;\displaystyle{#1}\;\;$}
}
\newcommand{\medbox}[1]{\fbox{%
\rule[-10pt]{0pt}{25pt}$\;\;\displaystyle{#1}\;\;$}%
}

\let\ssection=\section
\renewcommand{\section}{\setcounter{equation}{0}\ssection}

\def\besub{\begin{subequations}}
\def\esub{\end{subequations}}

\begin{document}
\preprint{\texttt{2004.02751v4 [gr-qc]}
}

\title{Geodesic motion in Bogoslovsky-Finsler Spacetimes
}

\author{
M. Elbistan$^{1}$\footnote{email:elbistan@impcas.ac.cn.},
P. M. Zhang$^{1}$\footnote{corresponding author. email:zhangpm5@mail.sysu.edu.cn},
N. Dimakis${}^{2}$\footnote{email:nsdimakis@gmail.com},
G. W. Gibbons$^{3}$\footnote{
email:G.W.Gibbons@damtp.cam.ac.uk},
P. A. Horvathy$^{4}$\footnote{email:horvathy@lmpt.univ-tours.fr}
}

\affiliation{
${}^1$ School of Physics and Astronomy, Sun Yat-sen University, Zhuhai, China
\\
${}^{2}$ Center for Theoretical Physics, College of Physical Science and Technology, Sichuan University, Chengdu 610065, China
\\
${}^3$D.A.M.T.P., Cambridge University, 
\\ Wilberforce Road,
\\ Cambridge CB3 0WA, United Kingdom
\\
${}^4$ Institut Denis Poisson CNRS/UMR 7013 - Universit\'e de Tours - Universit\'e d'Orl\'eans Parc de Grandmont, 37200, Tours, (France)}

\date{\today}


\begin{abstract}
We study the free motion of a massive particle moving in the background of a Finslerian deformation of a  plane gravitational wave in Einstein's general relativity. The deformation is a curved version of a one-parameter
family of  relativistic Finsler structures introduced by Bogoslovsky, which are invariant under a certain deformation of Cohen and Glashow's very special relativity group $\ISIM(2)$. The partially broken Carroll symmetry we derive using Baldwin-Jeffery-Rosen coordinates allows us to integrate the geodesics equations. The transverse coordinates of timelike Finsler-geodesics  are identical to those of the underlying plane gravitational wave for any value of the Bogoslovsky-Finsler parameter $b$. We then replace the underlying  plane gravitational wave with a homogenous pp-wave solution of the Einstein-Maxwell equations. We conclude by extending the theory to the  Finsler-Friedmann-Lemaitre model.
\\
\texttt{Phys. Rev. D 102, 024014}. DOI: 10.1103/PhysRevD.102.024014
\end{abstract}

\maketitle

\tableofcontents

\section{Introduction}\label{Intro}

Our present fundamental physical theories  are based
on local Lorentz invariance and hence on local isotropy.
This leads naturally to the introduction of pseudo-Riemannian
geometry and its associated metric tensor. It has long been known, however that a ``principle of relativity'' can be made compatible with anisotropy by deforming  the Lorentz group  by the inclusion of dilations \cite{Glashow} (although the experiments of Hughes, and  Drever indicate that the anisotropy must be very weak \cite{HughesD}).

Currently there is also a great deal of activity exploring the astrophysics and cosmology of alternative gravitational theories  based on standard Lorentzian geometry. Laboratory tests of local Lorentz invariance are very well developed and have reached impressive levels of precision.

Riemann himself envisaged more general geometries.
An elegant construction combining these ideas was provided some time ago by Bogoslovsky \cite{B1,B2} (for more recent accounts, see \cite{B4}).
In what is now known as Finsler geometry, the line element is a general homogeneous function of degree 1 in displacements, rather than the square root of a quadratic form.

The theory proposed by Bogoslovsky, which is the main subject of interest of this paper, has turned out to be relevant for  attempts to accommodate a proposal
of  Cohen and Glashow \cite{Glashow}, accounting for  weak CP violation in the standard model of particle physics, in the gravitational background \cite{Gibbons:2007iu,B3}.

The first significant application of Finsler geometry
to  physics is due to Randers \cite{Randers:1941gge} who pointed out that the world line
of  a particle of  mass $m$ and electric charge $e$ extremizes the action
\ben
S_0= \int\!\cL_0 d\lambda =- \int\!m\sqrt{-g_{\mu\nu} dx^\mu dx ^\nu} + e A_\mu dx^\mu\,,
\label{sqrtaction}
\een
where $\lambda$ is an arbitrary parameter and $A_\mu$ is the electromagnetic potential.
 Randers applied this idea to Kaluza-Klein theory. Further studies  followed \cite{Tavakol0,Tavakol,Roxburgh,RTVdB} ; it has also  been applied to the gravitomagnetic effects occurring in stationary spacetimes~\cite{Gibbons:2008zi}.
For more recent work on Finsler spaces, see \cite{Kostelecky11,Fuster1,Fuster2,Minguzzi,Perlick,Edwards,Bernal,Fuster20,Hohmann19,Caponio:2007dt,Caponio20}. Null geodesics and causality are considered in particular in
 \cite{Caponio:2007dt}.

The aim of the present paper is to contribute to the physical applications of the Finslerian generalization of general relativity by exploring the motion of freely moving massive particles in the background of Bogoslovsky-Finsler deformations of plane gravitational waves and spatially flat Friedmann-Lema\^{\i}tre cosmologies.

\section{Finsler spaces}\label{earlysec}

In this section, we shall briefly summarize
some earlier results  \cite{Tavakol0,Tavakol,Roxburgh,RTVdB}.
An excellent general reference to Finsler geometry used by these authors  is \cite{Rund} to which we refer the reader for more details of the general theory.
If $F(x^\mu, y^\nu)$ is a Finsler  function
\footnote{See Refs. \cite{Tavakol0,Tavakol,Roxburgh,Rund};
  $y^\mu$ is a four-velocity and $(x^\mu, y^\mu )$ are local coordinates on
$TM$, the tangent bundle of the spacetime manifold $M$.},
then it is assumed that $F^2 (x,y)$  may be written such that \cite{Roxburgh}
\ben
F^2(x,y) ={\cal F}(x,y) g_{\mu \nu} y^\mu y^\nu,
\label{one}
\een
where ${\cal F}(x,y)$  is a positive function which is positively homogeneous in the velocities.
Moreover \cite{Roxburgh},  if
$
H(x) _{\alpha_1,\alpha_2, \dots\, \alpha _N }
$
is a totally symmetric tensor of rank $N$
which is covariantly constant with respect to the Levi-Civita covariant derivative of the metric $g_{\mu \nu}$, then if
\ben
\omega = g(x)_{\mu \nu} y^\mu y^\nu /( H_{\alpha_1,\alpha_2, \dots, \alpha _N}
 y^{\alpha _1}  y^{\alpha_2 }\,\dots y^{\alpha _N} )^{2/N}
\aand
\cF(x,y) = \cF(\omega)
\een
then the set of Finsler geodesics of $F$ and the set of standard Riemannian geodesics of $g_{\mu \nu}$ coincide [see Eq. (20) in Ref. \cite{Roxburgh}.) Further aspects are considered in Refs. \cite{Hohmann19,Caponio20}.

The case of Finsler pp-waves \cite{Fuster1,Fuster2} occurs when $H$ is a covariantly constant null covector and
$g_{\mu \nu}$ is the metric of  a pp-wave, a special case of which is a plane gravitational wave. This was in effect  pointed out by  Tavakol and
Van der Bergh \cite{Tavakol} in 1986 and elaborated and extended by Roxburgh in 1991. Bogoslovky's original
flat Finsler metric \cite{B1,B2}
is a special case of  their work but no mention of Bogoslovsky is made in Refs. \cite{Tavakol0,Tavakol,Roxburgh}, and so one assumes that they were
unaware of it.

We next recall some  basic definitions and notation used in Refs. 
\cite{Tavakol0,Tavakol,Roxburgh}.
Given a Finsler function $F(x,y)$, one may define the \emph{Finsler metric tensor}
\ben
f_{\mu \nu}(x,y)  = \half \frac{\p^2 F^2 (x,y) }{\p y^\mu \p y^\nu }
\label{Fmetric}
\een
which is homogeneous of degree 0 in $y^\mu$.
That is, $ f_{\mu \nu}(x,y)$ depends only upon the \emph{direction}. Differentiating the identity
$ F^2 (x^\alpha, \lambda y^\mu)= \lambda ^2 F(x^\alpha, y^\mu)$ twice with respect to $\lambda$ implies
that
\ben
f_{\mu\nu}y^\mu y^\nu =F^2 (x,y) \,.
\een
The Finsler  line element or arc length $ds$  along a curve $\gamma$  with tangent vector
$y^\mu = \frac{dx^\mu}{d \lambda}$  is given by
\ben
ds ^2 = F^2(x^\mu,dx^\mu)  =  f_{\mu\nu}(x,y) dx ^\mu dx^\nu \,,
\een
and a Finsler  geodesic is one for which
$
\delta\! \displaystyle\int_\gamma\! F(x,dx^\mu) =\; \delta\!\! \displaystyle\int_\gamma\! ds  =0\,.
$
The Euler-Lagrange equations  are
\ben
\frac{d^2 x^\mu}{ds ^2} + \gamma ^\mu_{\alpha \beta}
\frac{dx^\alpha}{ds}\frac{dx^\beta}{ds} =0 \,,
\een
where
\ben
\gamma_{\mu\nu\kappa }= \half \bigl (\frac{\p f_{\kappa  \nu }}{\p x^\mu} +
  \frac{\p f_{\mu\kappa}}{\p x^\nu} - \frac{\p f_{ \mu \nu}}{\p x^\kappa } \bigr) \,, \qquad \gamma^\mu _{\nu \kappa}= f^{\mu  \sigma} \gamma_{\nu \sigma \kappa }
\een
are the analogues of \emph{Christoffel symbols of the first and second kind}, respectively.
In deriving the Euler-Lagrange equations, one
uses the fact that $y^\kappa \frac{\p f_{\alpha \beta}}{\p y^\kappa} =0$  because $f_{\mu \nu}$ is homogeneous of degree $0$ in $y^\mu$.
Evidently, under a change of parameter $s \rightarrow \lambda =\lambda(s)$ we have
$
\frac{d}{ds} = \lambda ^\prime  \frac{d}{d\lambda} \,,\, f_{\mu \nu} \rightarrow f_{\mu \nu}
$
since  $ f_{\mu\nu}$ is homogenous of degree 0 in velocities.
Thus, as in the standard Lorentzian situation,
\ben
\frac{d^2 x^\mu}{d \lambda^2} + \gamma ^\mu_{\alpha \beta}
\frac{dx ^\alpha}{d\lambda} \frac{dx^\beta}{d\lambda} = -
\frac{\lambda^{\prime \prime}}{\lambda ^\prime} \frac{dx^\mu}{d \lambda} \,.
\een
If $\lambda ^{\prime \prime}=0$, $\lambda$ is called an affine parameter, and in what follows, unless otherwise stated, $\lambda $ will denote an affine parameter.

In Refs. \cite{Tavakol0,Tavakol, Roxburgh} the quantities
\ben
G^\mu  =\half \gamma ^\mu \,_{\nu \kappa} y^\nu y^\kappa \,\qquad
G^\mu_{\nu \kappa} = \frac{\p^2 G^\mu}{\p y^\nu \p y^\kappa}
\een
are introduced. Although in general
$
G^\mu _{\nu \kappa} \ne \gamma^\mu _{\nu \kappa} \,,
$
by virtue of the  homogeneity of degree 0 of $\gamma^\mu_{\nu \kappa}$ in $y^\mu$,
one has
\ben
 G^\mu_{\nu \kappa}y^\nu y^\kappa
 = \gamma ^\mu_{\nu\kappa} y^\nu y^\kappa \,,
 \een
 and therefore Euler-Lagrangian equations may be rewritten as
\ben
\frac{d^2 x^\mu }{d \lambda ^2} + G^\mu _{\nu \kappa}
\frac{d x^\nu}{d \lambda}\frac{dx ^\kappa}{d \lambda }= 0.
\een
In general $ G^\mu \,_{\nu \kappa}$ depends upon the direction
 \cite{Tavakol} ; a
 \emph{Berwald-Finsler manifold}  is one for which
$ G^\mu \,_{\nu \kappa}$
is independent of the direction, --i.e.,
\ben
  G^\mu_{\nu\kappa} =  G^\mu_{\nu\kappa} (x)\,.
\een

The motivation for Refs. \cite{Tavakol0,Tavakol,Roxburgh} came from a classic paper of Ehlers, Pirani, and Schild  \cite{Ehlers} examining the fundamental assumptions justifying
the use of pseudo-Riemannian geometry adapted in Einstein's
general relativity. Roughly speaking the idea was that

\begin{itemize}

\item  {\it The Principle of  Universality of free fall}
  endows spacetime $\cM$  with a \emph{projective
  structure}, that is, an equivalence class of curves,
$\gamma : \lambda \in \mathbb{R} \rightarrow \cM $
 up to reparametrization.

\item {\it The Principle of  Einstein Causality} endows
spacetime with a causal structure such that light rays
are  determined  by some connection.

\end{itemize}

They conjectured that the only way of
achieving  this was that freely falling particles
and null rays
follow the geodesics of a pseudo-Riemannian  metric,
and in the case of particles that
the curves carry a privileged parametrization
given by proper time with respect to the pseudo-Riemannian metric
along their paths in specetime, whether freely falling or not.

In Refs. \cite{Tavakol0,Tavakol} Tavakol and Van den Bergh
sought to show that one could  pass to a Finsler structure as well,
provided one assumes

\begin{itemize}
\item \{Ai\}
\ben
F^2(x,y) = e^{2 \sigma(x,y)} g_{\mu \nu}(x)y^\mu y^\nu  \label{two}  \,,
\een
where $g_{\mu \nu}(x)$ is a Lorentzian metric and
$\sigma(x^\alpha, y^\mu)$ is homogeneous degree 0 in $y^\mu$.

This condition ensures that the conformal structures of the Finsler metric
and the Lorentzian metric agree locally, in the spirit  of
Ref. \cite{Ehlers}. It is pointed out in Ref. \cite{Tavakol} that Eq. (\ref{two})
is not equivalent to
\ben
f_{\mu \nu} =  e^{2 \sigma(x,y)} g_{\mu \nu}
\een
because it this were true  then $\sigma$ would only  depend upon $x$ and hence
$f_{\mu \nu}$  and $g_{\mu \nu}$ would be conformally  related, citing Ref. \cite{Rund}. \footnote{In fact, Eq. (\ref{two}) is obviously equivalent to Eq. (\ref{one})
which is the form used by Ref. \cite{Roxburgh} (who  appears  to regard it as always true, although $g_{\mu \nu}$ is not necessarily unique.}

\item \{Aii\}
 \ben
 G^\mu _{\nu \kappa} = \Bigl  \{{}^\mu _{\nu \kappa}  \Bigr\}
\een
where  $\Big \{{}^\mu _{\nu \kappa}  \Bigr \}  $ are the Christoffel symbols of the
 Lorentzian metric $g_{\mu \nu}$.
This condition ensures that the projective structures of
the Finsler structure $F(x,y)$ and the Lorentzian structure $g_{\mu \nu}$ agree locally, again in the spirit of Ref. \cite{Ehlers}.

\end{itemize}

Tavakol and Van den Bergh \cite{Tavakol} claimed that
the necessary and sufficient condition on $\sigma(x,y)$
is
\ben
\frac{\p \sigma}{\p x^\mu} - y^\nu \frac{\p \sigma}{\p y^ \kappa }
\Bigl \{{}^\kappa _ {\mu\nu}\Bigr\} =0\,.
\een
and referred to it as the \emph{metricity condition}.
The name originates in the theory of the so-called \emph{Cartan connection}. One defines
\ben
C_{\mu \nu \kappa} = \half \frac{\p f_{\mu \nu}}{\p y^\kappa}
\een
which is from Eq. (\ref{Fmetric}), as  totally symmetric in $\mu,\nu ,\kappa $.
Then one defines
\ben
\Gamma_{\mu \nu \kappa}= \gamma_{\mu\nu\kappa} -
\bigl( C_{\sigma\kappa\nu} \frac{\p G^\sigma}{\p y^\mu}
  + C_{\sigma \kappa \mu } \frac{\p G^\sigma }{\p y^\nu }   - C_{\mu \sigma \nu } \frac{\p G^\sigma }{\p y^\kappa}  \bigr)\,.
\een

Acting on a vector $W^\mu(x,y)$ the Cartan covariant derivative is defined by
\ben
\nabla^{\rm Cartan} , _\kappa\, W^\mu = \frac{\p W^\mu}{\p x^\kappa} +
- y^\sigma \Gamma ^\lambda _{\sigma \kappa} \frac{\p W^\mu} {\p y^\kappa}
+ \Gamma ^\mu_{\nu \sigma} W^\sigma
\een
and extended to tensors of arbitrary valence in the obvious way. The Cartan connection satisfies
\ben
\nabla ^{\rm Cartan} _\kappa \, f_{\mu \nu }= 0\,.
\een
This is equivalent to
\ben
\frac{\p F^2}{\p x^\kappa} - \frac{\p F^2 } {\p y^\sigma} \frac{ \p G ^\sigma}
{\p y^\kappa} \label{metricity} =0 \een
and ensures that the norms of the vector remain constant under parallel transport along different routes.
In Ref. \cite{Roxburgh} it is written as
  \ben
\frac{\p F}{\p x^\kappa} - \frac{\p F} {\p y^\sigma} \frac{ \p G ^\sigma}
{\p y^\kappa} =0 \,.
\een

 In Ref. \cite{Tavakol} it was suggested that
 \ben
g_{\mu \nu} x^\mu dx ^\nu  = -2 dudv + \alpha(u) dx^2 + \beta (u) dy^2
  \een
with $(x^1,x^2,x^3,x^4)=(x,y,u,v)$, which they call a plane wave, might lead to a solution. They find [in their Eq. (34)] that
\ben
\sigma = \sigma(\frac{\alpha \dot x^2 + \beta\dot y^2 -2 \dot u \dot v}{\dot u^2})
  \een
 and they claim that it is  indeed a solution.

The treatment of Ref. \cite{Roxburgh} starts
 with the helpful observation that
 the sums, products, and ratios of solutions are again
 solutions. Roxburgh investigated Lorentzian metrics with covariantly constant
vector fields and pointed  out that pp-waves are a special case.

\section{\B-F metrics}\label{BFsec}

Bogoslovsky's  theory \cite{B1,B2} was  based on the Finsler line element
such that the proper time $\tau$ along a future-directed  \emph{timelike}
world line $x^\mu(\tau)$ in flat  Minkowski
spacetime is obtained by combining the  Minkowski line element with
what we call here the Bogoslovsky factor,
\ben
d \tau=
(-\eta_{\mu\nu} dx^\mu dx^\nu )^{\frac{1-b}{2}}\,(-\eta _{\mu \nu}l^\mu d x^\nu)^b\,,
\label{Bogflat}
\een
where   $0\leq b<1$ is a dimensionless constant,      $\eta_{\mu\nu}$ is
the flat Minkowski metric tensor (with mainly positive signature), and
$l^\mu$, is a constant future directed null vector -- i.e.
\ben
\p_\mu l^\nu =0 \,,  \quad \eta _{\mu \nu}l^\mu l^\nu =0\,, \quad l^0 >0 \,.
\een
where $\p_\mu = \frac{\p}{\p x^\mu}$.
The Bogoslovsky factor makes \eqref{Bogflat} homogenous of degree 1 -- i.e., Finslerian.

The parameter $b$ introduces spatial anisotropy which might be relevant at the early stages of the Universe \cite{BogoGo}.
The constant $b$ is very small by Hughes-Drever-type experiments \cite{HughesD}; Bogoslovsky argues that $b<10^{-10}$ \cite{Bogo007}. For $b=0$ we recover the Minkowski proper time element, cf. Eq. \eqref{sqrtaction} with $A_\mu=0$.

Bogoslovsky's Finsler line element has an obvious generalization: in Eq. \eqref{Bogflat}, one
replaces $\eta_{\mu \nu} $  with a curved  pseudo-Riemannian metric $g_{\mu \nu}(x)$
 and $l^\mu$ with  a future-directed null vector  such that
\ben
\nabla_\mu l^\nu =0  \,, \qquad  g_{\mu \nu} l^\mu l^\nu=0
\,,\een
where $\nabla _\mu$ is the Levi-Civita connection of the  Lorentzian metric
$g_{\mu \nu}$.
This idea  has recently been explored
in Refs. \cite{Fuster1,Fuster2,Minguzzi,Bernal}, where  such spacetimes are  called ``Finsler pp-waves''.

Such spacetimes are also referred to as  Brinkman  \cite{Brinkmann} or
Bargmann \cite{Bargmann} spacetimes. They   admit Brinkmann coordinates
$X^\mu = (V,U,X^i)$  such that
\ben
g_{\mu\nu}dX^{\mu}dX^{\nu} = 2dVdU + dX^i dX^i - 2 H(X^i,U) d U^2\,,
  \label{pp}
\een
where the spacetime dimension is
$d+1$ and $i=1,2,\dots ,d-1$. $H(X^i,U)$  is an arbitrary not identically vanishing function of its arguments
\footnote {Our choice of sign for $g_{UV}$
has the advantage that raising and lowering of indices entails no minus signs, merely swapping $U$ and $V$ and  is consistent with our  previous papers. It has however the consequence  that if we choose a time orientation  such that   $U$ increases to the future then $V$ decreases to future.
 In other words $\frac{\p}{\p U}$ is a future directed null vector field and
  $\frac{\p}{\p V} $ is a past directed vector field.}.
$U,V$ may be written as
\beq
V= X^- =\frac{1}{\sqrt{2}}(X^{d} - X^0 ) \,,\; U = X^+=\frac{1}{\sqrt{2}} (X^{d} +  X^0).
\eeq
We have
$l^\mu \p_\mu = -  \p_V$\,
so that
$
-g_{\mu \nu}l^\mu dX^\nu = dU \,.
$
Then the Finsler pp-line element is
\ben
\bigbox{
d \tau=
(-g_{\mu \nu} dX^\mu dX^\nu)^{\frac{1-b}{2}}\,
(-g_{\mu \nu} l^\mu d X^\nu)^b
=
(-g_{\mu\nu} dX^\mu dX^\nu)^{\frac{1-b}{2}}\, (dU)^b\,,}
\label{Bogcurved}
\een
where $g_{\mu \nu}$ is the  pp-wave metric.

Returning to the pp-waves, we recall that
the metric is Ricci flat if and only if  $H(X^i, U)$ is  a harmonic function
of the coordinates $X^i$ ; it  may, however, have  arbitrary dependence upon $ U$.
It then represents a left-moving (i.e., in the   negative $X^{d}$  direction) gravitational wave  such that
$X^i$'s are transverse to the direction of motion. The  wave fronts
$U={\rm constant}$ are null hypersurfaces,  and the  covariantly
constant and hence Killing  null vector  field $\p _V$ lies in the wave fronts.

If, in addition, $H(X^i, U)$ is quadratic in the transverse coordinates
then we have a \emph{plane gravitational wave}. If $d=3$, which we assume from now on,
then
\ben
-2H= \cA_+(U) (X_1^2-X_2^2) +\cA_\times (U)  2X_1X_2 = K_{ij} (U) X^iX^j\,,
\label{HAAK}
\een
where $\cA_+(U)$ and  $\cA_\times(U)$ are  the amplitudes of the two
plane polarization states.

For general $\cA_+(U)$ and  $\cA_\times(U)$
there is a five-dimensional  isometry group $G_5$ which acts multiply transitively
on the three-dimensional wave fronts
$U={\rm constant}$ \cite{BoPiRo,exactsol}. This group is
a subgroup of the six-dimensional Carroll group
$\Carr (2)$ in three spacetime dimensions \cite{Leblond,Carrollvs} in which the $SO(2)$
subgroup is omitted \cite{Carr4GW}.

The  Carroll group $\Carr(2)$
 may be regarded as a   subgroup of the Poincar\'e group $ISO(3,1)$ defined by freezing out $U$-translations \cite{Carrollvs}; it
acts on the null  hyperplanes $U= {\rm constant} $.
If we label the Killing vector fields
of the Poincar\'e group as
\ben
P_\mu  = \frac{\p}{\p X^\mu} \,, \qquad
L_ {\mu \nu} = X_\mu P_\nu - X_\nu P_\mu \,,
\een
then the Carroll group is generated by
\besub
\begin{align}
P_- &= \frac{\p}{\p V} \,,\quad P_i = \frac{\p}{\p X^i}\,,
\label{Carrolltrans}
\\[6pt]
L_{ij}=X_i P_j - X_j P_i\, ,\quad
L_{-i}&={X_-}P_i-X_iP_-
=
U P_i- X_i P_- \,,
\label{Carrollrotboost}
\end{align}
\label{fullCarroll}
\esub
$i=1,2$, and each  hyperplane $U= {\rm \const} $, is left invariant. The generators in Eq. \eqref{Carrolltrans} are translations, whereas  those in Eq. \eqref{Carrollrotboost} are planar rotation and  boosts.
Since $d=3$, we may  relabel the generators $L_{ij}= J $, and the U-V boost
\beq
N_0 = L_{+-} = X_{+}P_- - X_{-}P_+
 =  VP_- -U_-P_+
 \where P_+=\frac{\;\partial}{\partial U}
\label{UVboost}
\eeq
and find that the four generators $N_0, J, L_{-i}$ generate a group which is abstractly isomorphic to the group $\SIM(2)$, the group of similarities, that is dilations, rotations, and translations of the Euclidean plane $\Bbb {E}^2$.
$\SIM(2)$ is the largest proper subgroup of the Lorentz group
$\SO(3,1)$. Adjoining the generators  $P_+, P_-, P_i $ gives rise to the eight generators of  $\ISIM(2)$, which is a  subgroup of the Poincar\'e group.  The group $\ISIM(2)$  acts
multiply transitively on Minkowski spacetime ${\Bbb E}^{3,1}$ .

It was suggested by Cohen and Glashow \cite{Glashow}
that  $\ISIM(2)$,  which may be thought of as the subgroup
of $\ISO(3,1)$ leaving invariant a null direction, could explain
weak CP violation while being compatible with tests  of Lorentz invariance,
since it would rule out \emph{spurions}, that is, tensor vacuum expectation values.

In  Ref. \cite{Coley} it was pointed out that Ricci flat
pp-waves are \emph{strongly universal}. In particular,
they have nonvanishing scalar invariants constructed from the Riemann
tensor and as a consequence satisfy almost any set of covariant field equations.
Quantum corrections to the metric vanish. Thus
this  property may  be thought of as the analogue
for the proposed  curved Bogoslovsky-Finsler structures with  a Ricci flat metric $g_{\mu \nu}$ of Cohen and Glashow's ``no spurions'' condition.

In Ref. \cite{Gibbons:2007iu}, an attempt was made to
find a link with general relativity  in which Minkowski spacetime ${\Bbb E}^{3,1}$
may be regarded as the coset
$\ISO(3,1)/\SO(3,1)$. The only two deformations
of the Poincar\'e group  led to the two
de Sitter groups $\SO(4,1)$ and $\SO(3,2)$
for which translations  act in a noncommutative fashion
on the cosets' de Sitter spacetime
$dS_4 = \SO(4,1)/\SO(3,1) $
and anti-de Sitter spacetime $AdS_4 = \SO(3,2)/\SO(3,1)$.

They therefore investigated  the deformations
of $\ISIM(2)$ and found that there exists
a  family of deformations depending upon
two dimensionless  parameters $a$ and $b$ .
However for all $a$ and $b$ the  translations
$P_+,P_-,P_i$ failed to commute. In general, the
rotation $J$ became a noncompact generator unless $a=0$, leaving
$\DISIM_b(2)$ depending on a dimensionless parameter $b$.
They then observed  that this is precisely the symmetry of Bogoslovsky's Finsler metric [Eq. \eqref{Bogflat}].
For a review of these ideas and their relation to the much earlier work of Voigt \cite{Voigt}, the reader is directed to the recent review
in Ref. \cite{Chashchina:2016tey}. For a recent discussion of Bogoslovsky-Finsler deformations in the light of the ideas of Segal, see Ref. \cite{Silagadze}.

In a recent paper \cite{Fuster1}, the authors
have  shown, among other things,  that the Bogoslovsky-Finsler pp-waves enjoy the same universal properties with respect to
generalizations of the  Einstein equations to
Finsler-Einstein equations as  those in the
pseudo-Riemannian  case discussed in  Ref. \cite{Coley}.

\section{Geodesics}\label{Geodesics}

The geodesics of a Finsler metric with Finsler function
 $F(x^\mu, \dot x^\mu)$, where $F$ is homogeneous of degree 1 in
$\dot x^\mu $,  are extrema of
\ben
I= \int  F\big(x^\mu, \frac{dx^\mu} {d \lambda}\big) d\lambda \label{action}\,.
\een
In the case we are considering,  we  restrict our attention to  future-directed  timelike  curves  for which both
$ g_{\mu\nu}l^\mu \dot x^\mu $
and  $-g_{\mu \nu }  \dot x^\mu \dot x^\nu $ are strictly  positive in order to ensure  that $F$ is real.
For a particle of mass $m$, the action with respect to  Lagrangians  is
\ben
S_{b}= 
 -m \int\! F \,d\lambda ,
\label{action2}
\een
where $F$ is the \B-F line element [Eq. \eqref{Bogcurved}].
The integral is independent of the parameter $\lambda$.  Therefore
if  $p_\mu = \frac{\p (-mF)}{\p \dot x^\mu}$,
 then
$ 
\cH  = p_\mu \dot x^\mu  +mF
$ 
is a constant of the motion. This is indeed true, but because
$F(x^\mu, \dot x ^\mu)$ is homogeneous of degree 1 in
$\dot x^\mu $ one has
$
\dot x^\mu \frac{\p F}{\p \dot x^\mu}= F
$
and consequently the constant vanishes identically. Standard Riemannian
or Lorentzian metrics are, of course, a special case of this general fact.

Now we analyze the motion along  geodesics
in Bogoslovsky-Finsler plane gravitational waves adapting the discussion for the standard  Einstein case given in Refs. 
\cite{Carr4GW,Conf4GW}.
First, we find it convenient to pass to Baldwin-Jeffery-Rosen (BJR) coordinates $x^\mu =(v,u,x^i)$, defined by
\ben
X^i=P_{ij}x^j\,,\quad U=u\,, \quad  V=v - \frac{1}{4} \frac{da_{ij}}{du} x^i x ^j  ,
\label{xfromX}
\een
where
$
a \; {\equiv (a_{ij})} = P^t P ,
$
and the matrix $P$ satisfies the matrix Sturm-Liouville equation
\ben
\frac{d^2 P}{du^2}= K P \,, \quad P ^t \frac{dP}{du}= \frac{dP}{du}^t P\,.
\label{SLeq}
\een
where $K=(K_{ij})$ is the profile in Brinkmann coordinates; see Eq. \eqref{HAAK}.

In BJR coordinates we have
\ben
g_{\mu \nu} dx^\mu dx^\nu = 2du d v  +  a_{ij}(u) \dot x^i \dot x^j \,,\qquad
l^\mu \frac{\p }{\p x^\mu} =  -\frac{\p}{\p v} \,.
\label{ppinBJR}
\een
Here
$ \dot x^\mu = \frac{dx^\mu}{d\lambda}$\, where $\lambda$ is an arbitrary parameter.
The Lagrangian is  proportional to the  Bogoslovsky-Finsler function,
\ben
\medbox{\cL_{b} = -m F \where
F= \Big(-2\dot u \dot v - a_{ij}(u) \dot x^i \dot x^j \Big)^{\half(1-b)}\,(\dot u)^b
}
\label{BFfunction}
\een

For the curve to be timelike we must
have $\dot u \dot v <0$ and $\dot u > 0$.
Since the integral \eqref{action} is independent of the choice of the parameter $\lambda$,  we are entitled
to make the choice $ \lambda = u$ and extremize
\ben
\int  \bigl (- 2 \frac{dv}{du} - a_{ij} (u)\frac{dx^i}{du} \frac{dx^j}{du}  \bigr ) ^{\half(1-b)} \, du \,. \label{action2}
\een
With this choice of parametrization the
integrand of \eqref{action2} is now  no longer homogeneous
in the velocities $\frac{dv}{du}$ and $\frac{dx^i}{du}$
but because $a_{ij}$ depends on the ``time''   $u$,  there is no
conserved analogue of the quantity $\cH$. The symmetry aspects will be further investigated in  Sec.\ref{Carrollsec}.

Before analyzing  the general case we recall, for  later comparison, some aspects of the geodesics of a pp-wave described by the square-root  Lagrangian [Eq. \eqref{sqrtaction}].

\kikezd{Geodesics in a pp-wave}

Let us thus first consider  a pp-wave  written in BJR coordinates, whose geodesics are
described by  Eq. \eqref{sqrtaction}  with $A_\mu=0$:
\beq
\cL_0 = -m\sqrt{-g_{\mu\nu} \dot{x}^\mu \dot{x}^\nu} =
 - m\sqrt{-a_{ij}(u) \dot{x}^i \dot{x}^j - 2\dot{u}\dot{v}}\; .
 \label{cL0}
\eeq
The canonical momenta $p_\mu = \frac{\partial \cL_0}{\partial \dot{x}^\mu}$ are
\beq
\label{L0cm}
p_u = \frac{m\dot{v}}{\sqrt{-g_{\mu\nu} \dot{x}^\mu \dot{x}^\nu}},
\quad
p_i = \frac{m a_{ij}\dot{x}^j}{\sqrt{-g_{\mu\nu} \dot{x}^\mu \dot{x}^\nu}},
\quad
p_v = \frac{m\dot{u}}{\sqrt{-g_{\mu\nu} \dot{x}^\mu \dot{x}^\nu}}
\eeq
of which $p_i$ and $p_v$ are constants of the motion since $a_{ij} = a_{ij}(u)$.  For a $u$-dependent profile $p_u$ is not conserved, though.
The geodesic equations of motion are\vspace{-2mm}
\begin{subequations}
\label{pregeoeq}
\begin{align}
\ddot{u} &= \dot{u} \frac{d}{d\lambda} \ln\big(\sqrt{-g_{\mu\nu} \dot{x}^\mu \dot{x}^\nu} \big),
\\[4pt]
\ddot{x}^i + \dot{u}a^{ij}a'_{jk}\dot{x}^k &= \dot{x}^i  \frac{d}{d\lambda} \ln\big(\sqrt{-g_{\mu\nu} \dot{x}^\mu \dot{x}^\nu} \big),
\\[4pt]
\ddot{v} -\frac{1}{2} a'_{ij}\dot{x}^i \dot{x}^j& = \dot{v}  \frac{d}{d\lambda} \ln\big(\sqrt{-g_{\mu\nu} \dot{x}^\mu \dot{x}^\nu} \big),
\end{align}
\end{subequations}
where $a'_{ij}= \frac{d a_{ij}}{d u}$.
Using the first equation, the two remaining ones simplify to
\footnote{Choosing the affine parameter $\lambda=u$,
the rhs would vanish.}
\vspace{-2mm}
\begin{subequations}
\label{geoeq}
\begin{align}
\ddot{x}^i + \dot{u}a^{ij}a'_{jk}\dot{x}^k & =\dot{x}^i \frac{\ddot{u}}{\dot{u}}\,,
\\
\ddot{v} -\frac{1}{2} a'_{ij}\dot{x}^i \dot{x}^j& = \dot{v} \frac{\ddot{u}}{\dot{u}}\,.
\end{align}
\end{subequations}
An ingenious way to solve these equations is to use the conserved quantities.
We first define the constants of the motion by setting
\beq
P_i = \frac{p_i}{p_v} = \frac{a_{ij}\dot{x}^j}{\dot{u}}
\,.
\eeq
The resulting first-order differential equation for $x^i$ is solved at once as
\beq
x^i{(u)}= \;\;\, S^{ij}(u)  P_j + x^i_0\,,
\label{transmot}
\een
where $S \equiv S^{ij}$ is the Souriau matrix  \cite{Carr4GW}, defined by
\ben
\frac{dS(u)}{du}= a^{-1}(u)\, .
\label{SHmatrix}
\een
 $p_v$ in Eq. (\ref{L0cm})  provides us in turn with a first-order equation for $v$:
\beq
\label{geodv}
\dot{v} = -\frac{1}{2} a^{ij} P_i P_j \,\dot{u}- \frac{1}{2}
\mu_0^2 
\,\dot{u}
\;\where\;
\mu_0=\frac{m}{p_v}\,.
\eeq
This equation is then  solved as
\beq
\label{vgeos}
v = -\frac{1}{2} P_i P_j S^{ij}(u) - \frac{1}{2}
\mu_0^2
\, u
 + v_0.
\eeq
The transverse motion \eqref{transmot} is the same for all values of the mass $m$, which enters  only the $v$ motion \eqref{vgeos} by a shift which is linear in $u$ and proportional to the mass-quotient term $\mu_0$ in Eq. \eqref{geodv}, familiar from Ref. \cite{nonlocal}.

\kikezd{Finsler Geodesics}.

Let us now consider the \B-F Lagrangian $\cL_b$ in Eq. \eqref{BFfunction}.
Its canonical momenta are
\begin{subequations}
\begin{align}
p_u &= m (\dot u)^{b-1} \big(-2 \dot u \dot v - a_{ij} \dot x^i \dot x^j\big)^{-\smallover{1+b}/{2}} \big((1+b) \dot u \dot v + b a_{ij} \dot x^i \dot x^j\big)\,,
\\
p_i &= m(1-b)(a_{ij}\dot{x}^j) \dot{u}^b ( -2\dot{u}\dot{v}-a_{ij}\dot{x}^i\dot{x}^j)^{-\frac{1+b}{2}}\,,
\\
p_v &=m(1-b) \dot{u}^{b+1}\big(-2\dot{u}\dot{v}-a_{ij}\dot{x}^i\dot{x}^j\big)^{-\frac{1+b}{2}}.
\end{align}
\label{BFmomenta}
\end{subequations}
$p_i$ and $p_v$ are constants of the motion as before and
we have the dispersion relation in Eq. (18) of Ref. \cite{Gibbons:2007iu}:
\ben
p^2\equiv g^{\mu\nu} p_\mu p_\nu = -m^2 (1-b^2)\dot{u}^{2b}
\Big(-2\dot v-a_{ij}\dot x^i \dot x^j \Big)^{-b}.
\label{dispersion}
\een

The geodesic equations,\vspace{-3mm}
\begin{subequations}
\label{BFpregeoeq}
\begin{align}
(b+1) \ddot{u} &= \dot{u} \frac{d}{d\lambda} \ln\big(-2\dot{u}\dot{v}-a_{ij}\dot{x}^i\dot{x}^j\big)^{\frac{1+b}{2}},
\label{BFueq}
\\[4pt]
\ddot{x}^i + \dot{u}a^{ij}a'_{jk}\dot{x}^k + b \frac{\ddot{u}}{\dot{u}}\dot{x}^i&= \dot{x}^i  \frac{d}{d\lambda} \ln\big(-2\dot{u}\dot{v} -a_{ij}\dot{x}^i\dot{x}^j\big)^{\frac{1+b}{2}},
\\[5pt]
\ddot{v} +\frac{3b-1}{2(1+b)} a'_{ij}\dot{x}^i \dot{x}^j + \frac{2b}{\dot{u}(1+b)}a_{ij}\dot{x}^i\dot{x}^j & = \frac{1}{b+1} \Big(\dot{v}+ \frac{2b}{1+b} a_{ij}\dot{x}^i\dot{x}^j \Big) \Big(\frac{d}{d\lambda} \ln\big(-2\dot{u}\dot{v} -a_{ij}\dot{x}^i\dot{x}^j \big)^{\frac{1+b}{2}}\Big),\quad
\end{align}
\end{subequations}
reduce to Eq. (\ref{pregeoeq}) when $b=0$.

The remarkable fact is that using Eq. \eqref{BFueq}, the two remaining equations  become the \emph{same} [Eq. \eqref{geoeq}], as for the square root Lagrangian [Eq. \eqref{cL0}].

This does \emph{not} imply identical solutions, though, as seen by  solving the geodesics equations  along the same lines as before. Setting once again  $P_i = \frac{p_i}{p_v}$  provides us with the transverse motion
\beq
\label{BFxgeos}
\medbox{
x^i(\lambda) = S^{ij} (u(\lambda)) P_j + x^i_0
}
\eeq
which is again Eq.  (\ref{transmot}).
 Then, from Eq. (\ref{BFmomenta}b), we infer that
\beq
\label{BFvgeos}
\dot{v} = -\frac{1}{2} \left(a^{ij}P_i P_j + \mu_b^2
 \right)\dot{u},
\;\where\;
\mu_b=\Big(\frac{m}{p_v}(1-b) \Big)^{\frac{1}{1+b}}\,,
\eeq
whose integration yields
\beq
\label{BFvgeos}
\medbox{
v = - \frac{1}{2}\,S^{ij}(u) \, P_i P_j
- \frac{1}{2}\mu_b^2\, u
+ v_0\,.
}
\eeq
Let us observe that this  takes the same form as  for Eq. (\ref{vgeos}), however, with a new, $b$-dependent mass-quotient term, $\mu_b$. For $b=0$ the latter reduces to $\mu_0$, and the  massive Eq. (\ref{vgeos}) is recovered.

The family of pp-wave geodesics are given
by  \eqref{transmot}  and \eqref{vgeos} and are labeled by the constants of integration  $P_i,\, x_0^i, v_0$ and $\mu _0$.
The Finsler geodesics are given by Eqs. \eqref{BFxgeos} and
\eqref{BFvgeos} and are labeled  by the constants of integration $P_i,\, x_ 0^i,\, v_0$ and $\mu_b$.
It is clear that the two sets of geodesics are identical
up to a $b$-dependent relabeling of the last constant of integration.

In the massless case $m=0$  (photons)  that, the $b$-dependent term drops out  from Eq. \eqref{BFvgeos}.
Letting $b\to1$ turns off the mass-quotient term, $\mu_b\to0$, and  all geodesics behave as if they were massless, consistently with Eq. \eqref{dispersion}.
See Fig. \ref{Steph386} in  Sec.\ref{Example}
for an illustration.

Another way to see the surprising identity of the geodesics is to consider the  Euler-Lagrange equations
\begin{equation}
  E_\mu  = \frac{\partial \cL_0}{\partial x^\mu} - \frac{d}{d t} \left( \frac{\partial \cL_0}{\partial \dot{x}^\mu}\right)=0
 \quad \text{and} \quad
  \tilde{E}_\mu  = \frac{\partial \cL_b}{\partial x^\mu} - \frac{d}{d t} \left( \frac{\partial \cL_b}{\partial \dot{x}^\mu}\right)=0
\end{equation}
of the two Lagrangians $\cL_0$ and $\cL_b$
in Eqs. \eqref{cL0} and \eqref{BFfunction}, respectively.

Both systems can be described by three independent equations, since the following identities hold:
$ 
  \dot{x}^\mu E_\mu \equiv 0, \, \dot{x}^\mu \tilde{E}_\mu \equiv 0.
$ 
Then the combinations are as follows:
\benu
\item
For the first system :
\begin{align}
  & \frac{\left( - g_{\mu\nu} \dot{x}^\mu \dot{x}^\nu \right)^{1/2}}{m} \left(\frac{2\dot{v}}{\dot{u}} E_v + \frac{\dot{x}^i}{\dot{u}} E_i \right) =0 \\
  & \frac{\left( - g_{\mu\nu} \dot{x}^\mu \dot{x}^\nu \right)^{1/2}}{m} \left(\frac{\dot{x}^i}{\dot{u}}E_v - a^{ij} E_j\right) =0.
\end{align}

\item
For the second system: 
\begin{align}
  & \frac{\left( - g_{\mu\nu} \dot{x}^\mu \dot{x}^\nu \right)^{\frac{1+b}{2}}}{m} \left[\frac{1}{(1-b^2)\dot{u}^{b+2}} \left(2 \dot{u}\dot{v} - b a_{ij}\dot{x}^i \dot{x}^j\right) \tilde{E}_v + \frac{\dot{x}^i}{(1-b)\dot{u}^{b+1}} \tilde{E}_i \right]=0 \\
  & \frac{\left( - g_{\mu\nu} \dot{x}^\mu \dot{x}^\nu \right)^{\frac{1+b}{2}}}{m} \left[ \frac{\dot{x}^i}{(1-b)\dot{u}^{b+1}}\tilde{E}_v + \frac{a^{ij}}{(b-1) \dot{u}^b} \tilde{E}_j \right] =0
\end{align}
\eenu
(where $i,j=1,2$) yield both parts of Eq. \eqref{geoeq}.

\section{Partially broken Carroll symmetry}\label{Carrollsec}

Generic plane gravitational waves are invariant under the same five-parameter
group we denote by $G_5$ \cite{BoPiRo,exactsol}. Expressed in
BJR coordinates, $G_5$ is implemented as \cite{Carr4GW},
\ben
u \rightarrow u \,,\quad
\bx \rightarrow \bx + S(u) \bb + \bc \,,\quad
v \rightarrow v- \bb \cdot \bx - \half \bb \cdot S(u) \bb + f \,,
\label{Carrollaction}
\een
where $S$ is Souriau's  matrix [Eq. \eqref{SHmatrix}].
The 2-vectors  $\bb$ and $\bc$  and $f$ are constants, interpreted as boosts, and as transverse and vertical translations.
These same transformations are isometries also for the \B-F metric  [Eq. \eqref{Bogcurved}] because $u$ is fixed, and the pp isometries leave the pp-wave metric -- and hence their powers -- invariant.
The transformations in Eq. \eqref{Carrollaction}
are generated by the vector fields
\beq
B_i=S_{ij}(u)\p_j -x_i\p_v,\quad \p_i
\aand
\p_v\,,
\label{Yiso}
\eeq
respectively. The only nonvanishing Lie bracket is
\ben
[\p_i, B_j ] = - \delta_{ij} \p_v \,.
\een
Rotations, generated by $L_{ij}=x_i\p_j- x_j\p_i$, are not symmetries in general.

The restriction of a pp-wave to the $u=0$ hypersurface $\cC_0$ carries a Carroll structure. The ``vertical'' coordinate $v$ is interpreted as ``Carrollian time'' \cite{Leblond,Carrollvs,BGL}.
$\cC_0$ is left invariant by the action \eqref{Carrollaction} and the generators then satisfy the Carroll algebra in two space dimensions with rotations omitted \cite{Carr4GW}. Then Eq. \eqref{Carrollaction} tells us how the Carroll group is implemented on any hypersurface $u=u_0=\const$

In the flat case, $a_{ij}= \delta_{ij}$, we have further symmetries.
In particular, adding the vector fields
$ \p_u,\, L_{ij}$  and
$ L_{+-}=v\p_v-u\p_u$
yields the Lie algebra of an eight-parameter 
 subgroup of the Poincar\'e group. We now have
\ben
[v\p_v-u\p_u, \p_v] = -\p_v \,,
\een
and so the \emph{direction} of the null Killing vector field $\p_v$ is preserved.
The eight-dimensional group they generate is  $ \ISIM(2) $.  Omitting
the translations $\p_v, \p_u, \p_i $ gives $\SIM(2)$, the largest proper subgroup
of the Lorentz group $\SO(3,1)$. This is the symmetry of Cohen and Glashow's
\emph{very special relativity}  \cite{Glashow}.

Returning to the case  of general pp-waves and their \B-F version [Eq. \eqref{Bogcurved}],
we emphasize that the BJR matrix $a=(a_{ij})$ and thus the Souriau matrix $S$, depends on the pp-wave metric only, but \emph{not} on the deformation parameter $b$. Therefore, the isometries in Eq. \eqref{Carrollaction} act, for  Eq. \eqref{Bogcurved}, exactly  as for standard plane waves.

The invariance of the \B-F model can be  confirmed
with respect to the partially broken Carroll group.
The infinitesimal version of  Eq. \eqref{Carrollaction} is $Y_{iso}$ in Eq. \eqref{Yiso}.
The linear momenta in Eq. \eqref{BFmomenta} are readily recovered ;  using Eq. \eqref{Yiso} for boosts, we get in turn,
\beq
k^i=p_v x^i-S_{ij}p_j,
\label{boostmom}
\eeq
-- just as for a gravitational wave \cite{Carr4GW}. Its conservation follows from Noether's theorem, and can also be confirmed by a direct calculation.
The dependence on $b$ is hidden in the momenta in Eq. \eqref{BFmomenta}. The initial position $x_0^i$ in Eq. \eqref{transmot} is the conserved value of $k^i$.


For $b=0$ the flat Bogoslovsky-Finsler model has one more isometry, identified with the U-V boost $N_0=L_{+-}$ in Eq. \eqref{UVboost}. For $b\neq0$, this generator is broken but  not entirely lost~:
  Let us explain how this comes about.

As said above, the (rotationless) Carroll isometry group $G_5$  in Eq. \eqref{Carrollaction} of the initial pp-wave remains a symmetry with identical generators for its \B-F extension.

To see what happens to U-V boosts we start with the Minkowski metric, $\eta_{\mu\nu} dx^\mu dx^\nu = \delta_{ij}dx^idx^j+2dudv$. A U-V boost,  implemented as
\beq
u\to {\lambda}^{-1} u,\quad x^i \to x^i,\quad
v \to  {\lambda} v
\label{uvboost}
\eeq
where $\lambda=\const>0$
is an isometry.
Moreover, its $b$-dependent deformation of Eq. \eqref{uvboost},
\beq
u \to \lambda^{b-1} u,  \quad x^i \to \lambda^b x^i, \quad v\to \lambda^{b+1} v
\label{buv}
\eeq
is readily seen to leave the \B-F line element \eqref{Bogflat} invariant -- although for $b\neq0$ it is  only a conformal transformation for the Minkowski metric,
$\eta _{\mu\nu} dx^\mu dx^\nu \to \lambda^{2b}\eta_{\mu \nu} dx^\mu dx^\nu$, and {not} an isometry
\footnote{
Noting that
$\lambda^{b-1}=(\lambda^b)^{\frac{b-1}{b}}$ shows that \eqref{buv} has dynamical exponent
$
z= 1-\frac{1}{b}<0\,
$
which corresponds to the conformal Galilei algebra labeled by $z$  \cite{MaxSIM,ConfGal}.
}.
We record for later use that the $b$-deformed boost [Eq. \eqref{buv}] is generated by
\beq
N_{b}=(b-1) u \partial_u + (b+1) v \partial_v + b x^i \partial_i\,.
\label{Nb}
\eeq

Both Eqs. \eqref{uvboost} and \eqref{buv} leave the hypersurface $u=0$ invariant, and extend  the Carroll action \eqref{Carrollaction}. We note that the restriction to $\cC_0$ of the deformed U-V boost [Eq. \eqref{buv}] scales the Carrollian time, $v \to \lambda^{b+1} v$. Therefore it is only the  \emph{direction} of $\p_v$ (and not $\p_v$ itself) which is preserved the isometry [Eq. \eqref{Bogflat}] is ``chronoprojective''  \cite{5Chrono,Conf4GW}:
\beq
\p_v \to \lambda^{-1-b}\, \p_v\,.
\eeq

In the flat case,  two more isometries --- namely $u$-translations and rotations complete the algebra to  one with eight-parameter. With some abuse, we will still refer to $G_5$ extended with $U$-translations (but with no rotations) as ``Carroll'' for simplicity and denote it by $G_6$. Its further extension by $U-V$ boosts will be called chrono-Carroll \cite{Conf4GW} and denoted by $G_7$.

The Lie algebra structure is most easily checked by taking the commutators of the vector fields  in Eqs. \eqref{fullCarroll} and \eqref{UVboost} and comparing with those  given in Eq. (9) of Ref.  \cite{Gibbons:2007iu},
which gives the structure constants of the deformed group $\DISIM_b(2)$ ; those of  $\ISIM(2)$  are obtained by setting $b=0$.

Further insight is gained by decomposing the deformed U-V boost generator $N_{b}$ in Eq. \eqref{Nb}
 into the sum of the undeformed expression $N_0=L_{+-}$ and a relativistic dilation, $D$:
\ben
N_{b} =
v \partial_v -u \partial u  + b (u \p_u + v \p_v + x^i \p_i)  = N_0 + bD\,.
\label{Nbdecomp}
\een
For $b\neq 0$, $N_0$ is broken, and it is only the above combination of U-V boosts and dilations which is a symmetry --- a situation familiar from gravitational plane waves \cite{Carr4GW,Ilderton,Conf4GW}.

It is instructive to see how this comes about.
In the flat Minkowski case $a_{ij} = \delta_{ij}$ and
Eq. \eqref{BFmomenta}  yield
$$
p_i = (\dot x_i /\dot u) p_v
\aand
p_u = \dfrac {(1+b)\dot u \dot v - b \dot x^i \dot x_i}{(1-b)\dot u^2} p_v\,.
$$
Then for  $\cD=D^\mu p_\mu$   and $\cN_0=N^\mu_0 p_\mu$ we have
\ben
\dot{\cD} =\dfrac {(2 \dot u \dot v + \dot x^i \dot x_i)}{(1-b)\dot u}\, p_v
\aand
\dot{\cN}_0 =  -b \dfrac {(2 \dot u \dot v + \dot x^i \dot x_i)}{(1-b)\dot u}\, p_v
= -b \dot{\cD}\,,
\een
so that the combination of the two  expressions  is  conserved~:
\beq
\dot{\cN}_b = 0
\qquad\text{for}\qquad \cN_b= \cN_0 + b \cD\,.
\label{buvcN}
\eeq

Now we turn to the curved case. Let us consider  a conformal transformation $f$ of a pp wave with metric $g_{\mu\nu}$:
\ben
f_\star g_{\mu \nu} = \Omega^2 g_{\mu \nu} \,,  \label{condition1}
\een
where $f_\star$ is the pullback map. This changes the ``pp factor'' in Eq. (\ref{ppinBJR}), as
\beq
  (g_{\mu \nu}dx^\mu dx^\nu)^{\half(1-b)} \to \Omega^{1-b}  (g_{\mu \nu}dx^\mu dx^\nu)^{\half(1-b)}.
 \eeq
 The change can be compensated by the  ``B-F factor'', though. Assuming that
$
f_\star l_\mu =  \Omega^a l_\mu\,
$
for some constant $a$ yields
\beq
 (l_\mu dx^\mu)^b (g_{\mu\nu}dx^\mu dx^\nu )^{\half(1-b)} \to \Omega^{ab+1-b}(l_\mu dx^\mu)^b (g_{\mu \nu}dx^\mu dx^\nu )^{\half(1-b)}.
 \label{actiononBF}
 \eeq

In the undeformed case $b=0$ this is a conformal transformation with conformal factor $\Omega$; obtaining an isometry
 requires $f$ to be an isometry for the pp-wave. This is consistent with our findings in the flat case for the U-V boost [Eq. \eqref{uvboost}].

For $b\neq0$ we have another option. If the exponent of $\Omega$ in Eq. \eqref{actiononBF} vanishes,
\beq
\label{condition2}
ab = b-1,
\eeq
then we do get an isometry again. For the $b$-deformed U-V boost in Eq. \eqref{buv}, we have $\Omega=\lambda^b$,  consistently with Eq. \eqref{condition2}.

In the case of plane gravitational waves,
one drops the angular momentum $J$ and $ \p_U$  and then the issue is what to do about $N_b$.
Our only certainty so far is that the flat-space implementation [Eq. \eqref{Nb}] does not work.

The symmetry of the  \B-F model is in fact of the very special relativity (VSR) type --- more precisely, a subgroup of the eight-parameter $\DISIM_b(2)$,  where $0<b<1$ is a deformation parameter \cite{MaxSIM,B3}.
$\DISIM_b(2)$ is isomorphic to the conformal Galilei group with dynamical exponent \cite{ConfGal}
\beq
z=1-\frac{1}{b}\,.
\label{dynz}
\eeq
For $b\neq0$, U-V boosts (which are isometries for the Minkowski case) are deformed to Eq. \eqref{Nb}, a  combination  of U-V boosts and relativistic dilations.

One can be puzzled about whether the ``deformation trick'' can work  also for a nontrivial profile.
The answer is that it \emph{might} work for a particular profile. Let us consider, for example, a pp wave [Eq. \eqref{pp}] written in Brinkmann coordinates with the
 (singular) profile
\beq
2H(X^i,U)=
-\frac{K^0_{ij}}{U^2}\,X^i X^j .
\quad K^0_{ij}=\const
\label{Ilderprof}
\eeq
This wave has a six-parameter isometry group \cite{Ilderton,AndPrenc,exactsol,Conf4GW}.
It is, in particular, invariant under a U-V boost  [Eq. \eqref{uvboost}].
 Then we find that the deformed U-V boost,
\beq
U \to \Lambda U, \quad \bm{X} \to \Lambda^{\frac{b}{b-1}}\bm{X}, \quad V \to \Lambda^{\frac{b+1}{b-1}} V
\label{modUVboost}
\eeq
leaves the \B-F line element
\beq
ds_{BF} = \Big(-2dUdV -d\bm{X}^2 - \frac{K^0_{ij}}{U^2} X^i X^j dU^2\Big)^{\frac{1-b}{2}}\big(dU\big)^b
\eeq
invariant. The usual U-V boost is recovered for   $b=0$.
Writing $\lambda=\Lambda{\frac{b}{b-1}}$ shows, moreover, that when $b\neq0$, the dynamical exponent is  $z=-1+\frac{1}{b}$, minus that in Eq. \eqref{dynz}.
Note that Eq. \eqref{modUVboost} is, once again, a conformal transformation of the pp-wave metric [Eqs. \eqref{pp}-\eqref{Ilderprof}] with conformal factor $\Omega^2=\Lambda^{\frac{2b}{b-1}}$.


\goodbreak
\section{Prolongation vectors and symmetries}\label{Prolongsec}

The connection of the aforementioned symmetries to integrals of the motion is established through Noether's first theorem \cite{Noether}~: each generator of any finite-dimensional Lie group of transformations which leaves the action form-invariant up to a surface term \cite{BH} is associated with a conserved quantity.

Consider, for example,  a dynamical system with dependent and independent variables $x^\mu(\lambda)$ and $\lambda$, respectively. The most general point transformation one can have is,
\begin{equation}
\label{genvf}
  \Upsilon = \sigma(\lambda, x)\frac{\partial}{\partial \lambda} + Y^\mu(\lambda, x) \frac{\partial}{\partial x^\mu}\,,
\end{equation}
where the coefficient $\sigma(\lambda, x)$ accounts for  transformations which might also involve the parameter $\lambda$. This vector can be extended to the space of the first derivatives $\dot{x}^\mu= \frac{d x^\mu}{d\lambda}$, i.e., we can consider the \emph{first prolongation} of $\Upsilon$,   defined as \cite{Olver,Christo}
\begin{equation}
\label{progen}
  pr^{(1)} \Upsilon = \Upsilon + \Phi^\mu \frac{\partial}{\partial \dot{x}^\mu}, \quad \text{where} \quad \Phi^\mu = \frac{d Y^\mu}{d\lambda} -\dot{x}^\mu \frac{d\sigma}{d\lambda}.
\end{equation}

The coefficient $\Phi^\mu$ here is to guarantee the correct transformation law
for the derivatives. Given, for example, the generator \eqref{genvf},  up to first order in the  transformation parameter (say $\epsilon$) we may write
\ben
  \bar{\lambda}
   \sim \lambda + \epsilon\, \sigma(\lambda,x)
\qquad
  \bar{x}^\mu
\sim x^\mu + \epsilon Y^\mu(\lambda,x)
\een
which furthermore implies
\begin{equation}
  \frac{d \bar{x}^\mu}{d\bar{\lambda}} \sim \frac{d(x^\mu+ \epsilon Y^\mu)}{d (\lambda + \epsilon \, \sigma)} \sim \left( \frac{d x^\mu}{d\lambda}+\epsilon \frac{d Y^\mu}{d\lambda}  \right)\left(1 - \epsilon \frac{d \sigma}{d\lambda} \right) \simeq \dot{x}^\mu + \epsilon \Phi^\mu .
\end{equation}

With the use of the extended vector $pr^{(1)} \Upsilon$, the initial requirement of Noether's theorem written as  \ben
\delta(L d\lambda)= d \Sigma
\label{deltaL}
\een
where $\Sigma=\Sigma(\lambda,x)$ is some function  can be cast in infinitesimal form as
\begin{equation}
  pr^{(1)} \Upsilon (L) + L \frac{d\sigma}{d\lambda} = \frac{d\Sigma}{d \lambda}.
\label{genprol}
\end{equation}
To an appropriate generator $\Upsilon$ and a function $\Sigma$ satisfying the above relation there corresponds a conserved quantity :
\begin{equation}
\label{intofmo}
  I = Y^\mu \frac{\partial L}{\partial \dot{x}^\mu} - \sigma (\dot{x}^\alpha \frac{\partial L}{\partial \dot{x}^\alpha}-L) - \Sigma \,.
\end{equation}

The geodesic system is invariant under arbitrary changes of the parameter $\lambda$; therefore  the inclusion of the coefficient $\sigma$ into Eq. \eqref{genvf} does not  contribute in the conservation law. As  can be seen using Eq. \eqref{intofmo}, $\sigma$ essentially multiplies the Hamiltonian, which is identically zero for Lagrangians which are homogeneous functions of degree 1 in the velocities.
The coefficient $\sigma$ plays a  role instead in \emph{Noether's second theorem} and an identity among the Euler-Lagrange equations of motion \cite{Sund}. As a result, we may restrict ourselves to consider pure spacetime transformations generated by vectors $Y=Y^\alpha (x) \partial_\alpha$. Then the first prolongation becomes
\begin{equation}
  pr^{(1)}Y = Y + \frac{d Y^\alpha}{d\lambda}\frac{\partial}{\partial \dot{x}^\alpha} = Y^\alpha(x) \frac{\partial}{\partial x^\alpha} + \frac{\partial Y^\alpha}{\partial x^\beta}\dot{x}^\beta \frac{\partial}{\partial \dot{x}^\alpha}\,.
\end{equation}
and Eqs. \eqref{genprol} and \eqref{intofmo} reduce to
\ben
 pr^{(1)} Y (L) = \frac{d\Sigma}{d \lambda}\,,
 \qquad
 I = Y^\mu \frac{\partial L}{\partial \dot{x}^\mu} - \Sigma\,.
\label{Noethercharge}
\een

If for a given spacetime vector $Y$, the relation $pr^{(1)} Y (L)=0$ is satisfied (as for isometries of the geodesic system), then $\Sigma$ is just a constant and can be omitted, thus having $\tilde{I}=I+\Sigma = Y^\mu \frac{\partial L}{\partial \dot{x}^\mu}= \const$

To illustrate the prolongation technique,
  we note that for a system in the background  $g_{\mu\nu}$  [Eq. \eqref{ppinBJR}] whose Lagrangian is $L$, the first prolongation of the isometries  in Eq. \eqref{Yiso},
 $$
  Y_{iso}=(S^{ij}\beta_j+\gamma^i)\,\partial_i + (-\beta_i x^i+\varphi)\,\partial_v\,,
$$
is
\beq
\label{prolong}
pr^{(1)}Y_{iso} (L) = \left(Y^\alpha_{iso} \partial_\alpha + \frac{\partial Y^\alpha_{iso}}{ \partial x^\beta}\ \dot{x}^\beta \frac{\partial}{\partial \dot{x}^\alpha}\right)(L).
\eeq
If the rhs  is a total derivative, then we have a symmetry for the system.

Applying Eq. (\ref{prolong}) first to the pp-wave Lagrangian
\beq
L_{pp} = \dot{u}\dot{v}+ \frac{1}{2} a_{ij}\dot{x}^i \dot{x}^j
\label{Lpp}
\eeq
 confirms that $Y_{iso}$ is a symmetry for the pp-wave.

Next, for the \B-F Lagrangian $\cL_{b}$ in Eq. \eqref{BFfunction},  we find that the rhs of Eq. (\ref{prolong}) vanishes:
\beq
pr^{(1)}Y_{iso} (\cL_{b}) = \left(m (1-b) (-a_{ij}\dot{x}^i\dot{x}^j -2\dot{u}\dot{v})^{-\frac{1+b}{2}}\dot{u}^b\right)\ pr^{(1)}Y_{iso} (L_{pp}) =0,
\eeq
proving that the Carroll group (with broken rotations) generates symmetries also for the \B-F metric. The conserved quantities listed in Sec.\ref{Carrollsec} are  recovered using Eq. \eqref{Noethercharge}.

Turning now to U-V boosts
 we check first that for the \emph{flat Minkowski metric} the prolongation  of the deformed boost $N_b$ in Eq. \eqref{Nb} vanishes:
\beq
pr^{(1)}N_{b}(\cL_0)
= 0
\label{Nb2}
\eeq
and thus generates the constant of the motion $\cN_b$ in Eq. \eqref{buvcN}.

However, the same calculation carried out in the \emph{curved} background $g_{\mu\nu}$ [Eq. \eqref{ppinBJR}] yields instead
\beq
pr^{(1)} N(\cL_{b}) = m u\ \dot{u}^b (b-1)^2 \left(\frac{d a_{ij}}{du} \dot{x}^i \dot{x}^j \right)(-a_{ij}\dot{x}^i \dot{x}^j -2 \dot{u}\dot{v})^{\frac{-1-b}{2}}\,.
\eeq
Consistently with what we said before, this vanishes  for the flat metric $\eta_{\mu\nu}$. However, it
 is manifestly \emph{not} a total derivative in general whenever $a=(a_{ij})$ is not a constant matrix.

\section{An Einstein-Maxwell example}\label{Example}

In this section, we treat the motion in the Bogoslovsky-Finsler deformation of a pp-wave which is not Ricci flat. It  is
\beq
 ds^2 = (dX^1)^2 + (dX^2)^2 + 2dUdV-\frac{\omega^2}{4}\Big((X^1)^2 +(X^2)^2\Big)dU^2\,.
\label{Brdmetric}
 \eeq
From Eq. (24.5) on p. 385 of  Ref. \cite{exactsol}, one learns that it  belongs to a class first considered by Baldwin and Jeffery \cite{Baldwin}. It is   conformally flat and is an Einstein-Maxwell solution with a covariantly constant null Maxwell field.
From the Bargmann point of view, this metric describes an isotropic harmonic oscillator in the plane with frequency $\omega$ \cite{Bargmann}. The kinematic group arising from  the null reduction is the Newton-Hooke group
 \cite{Gibbons:2003rv}.  Because the metric \eqref{Brdmetric} is of the form of Eq. \eqref{pp} with
\ben
-2H= K_{ij} X^iX^j
\een
where $K_{ij}$ is nondegenerate and independent of $U$, it is also a Cahen-Wallach symmetric space
\cite{Cahen,Figueroa-OFarrill:2001hal,
Blau:2002js,Blau:2003ia}.
Following the procedure outlined in Sec.\ref{Geodesics}, the metric  \eqref{Brdmetric} can presented in the BJR form. We set $a=P^tP$, where
\beq
P =
\left[\begin{array}{rrr}
\Big(1-\sin({\omega}u)\Big)^{1/2} \cos\phi && -\Big(1+\sin({\omega}u)\Big)^{1/2} \sin\phi
\\[8pt]
\Big(1-\sin({\omega}u)\Big)^{1/2} \sin\phi  && \Big(1+\sin({\omega}u)\Big)^{1/2} \cos\phi
\end{array}\right]
\eeq
is a solution of the Sturm-Liouville equation
\eqref{SLeq} with  diagonal profile $K=-\frac{\omega^2}{4}\Id$. Then, using Eq. \eqref{xfromX},  we end up with
\ben
ds^2 = \big(1-\sin(\omega u)\big) dx^2 + \big(1+\sin (\omega u)\big) dy^2 + 2 du dv
\label{BrdBJR}
\een
which has $a = P^T P = \diag\big(
1-\sin(\omega u), 1+\sin(\omega u)\big)
$.
On p. 386 of Ref. \cite{exactsol}, this result is ascribed to Brdi\u{c}ka \cite{Brdicka}.
Equation \eqref{BrdBJR} shows that the U-V boost symmetry is manifestly broken.

\begin{figure}[h]
\includegraphics[scale=.45]{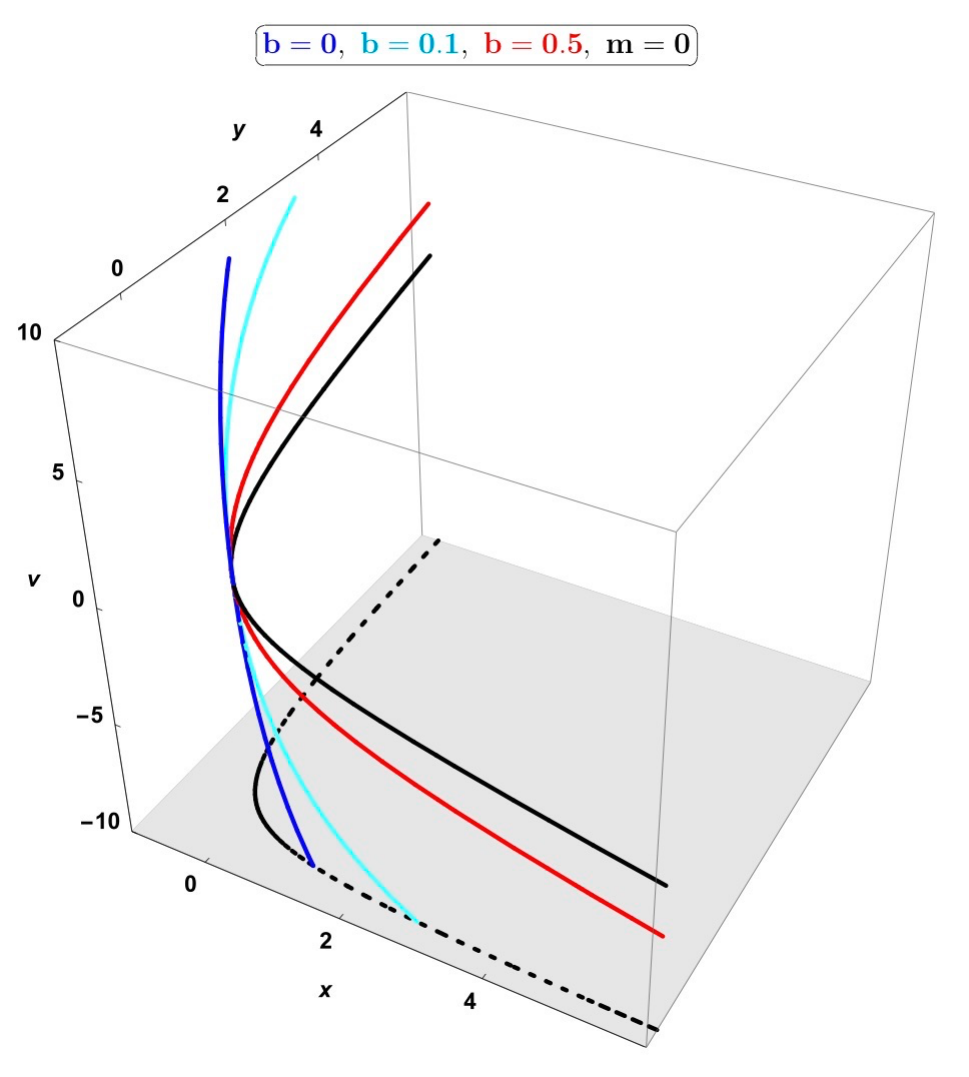}
\vskip-5mm
\caption{\textit{\small Consistently with \eqref{transmot}, the  Bogoslovsky-Finsler  geodesics project  to the same curve in 2D transverse space for all values of the parameter $b$ while their $v$ coordinates differ, according to \eqref{BFvgeos}, in a $b$-dependent term, which is linear in retarded time, $u$. Experiments indicate that the anisotropy and hence $b$ is  very small. When $b\to1$ the trajectory approaches to the massless one (in heavy black), consistently with \eqref{BFvgeos}.
\label{Steph386}
}}
\end{figure}

The Souriau matrix is
 found by integrating the inverse of $(a_{ij})$, cf. Eq. \eqref{SHmatrix},
\beq
S(u) = \frac{1}{\omega}
\left[\begin{array}{cc}
\tan\left(\frac{\omega u}{2}+\frac{\pi}{4}\right) + C_1& 0
\\[6pt]
0 & \tan\left(\frac{\omega u}{2}-\frac{\pi}{4}\right) + C_2
\end{array}\right]\, ,
\eeq
where $C_{1,2}$ are integration constants. Choosing $u_0 = 0$ yields $C_1 = -1$ and $C_2 = 1$.
The trajectories \eqref{transmot}-\eqref{BFvgeos} for different values of $b$ are depicted in Fig.\ref{Steph386}.

We mention that the profile of the metric \eqref{Brdmetric}  is $U$-independent and therefore  $U$-translation, $U\to U+\epsilon$ is an additional isometry. This carries over trivially to its Finslerized line element [Eq. \eqref{Bogcurved}], since both  the pp-wave metric and the ``\B-F factor" are  invariant.

\vskip-3mm

\section{Bogoslovsky-Finsler-Friedmann-Lema\^{\i}tre  Model}\label{FFLsec}

In this section we shall describe a simple extension
of Bogoslovsky's theory to take into account the expansion of the Universe. For some previous work, see Refs.
\cite{Roxburgh2,Bekenstein:1992pj,  Kouretsis:2008ha,Hohmann:2016pyt}.
In contrast to our work, these authors  consider only Finsler metrics which share the isotropy and spatial homogeneity of Friedmann-Lema\^{\i}tre models. This  necessarily excludes the use of a null vector field.

\subsection{The $\Lambda$CDM model}
The simplest standard model consistent with current observational data is the
spatially flat Friedmann-Lema\^{\i}tre  model with metric
\ben
g_{\mu \nu} dx^\mu dx^\nu = -dt^2 + a^2 d\bx ^2
\label{FLmetric}
\een
where $a=a(t)$ and $\bx =(x,y,z)$. The ``scale factor'' $a(t)$ is determined
by the Einstein equations once the matter content has been specified.
The favored $\Lambda$CDM model has
\ben
a(t) = \sinh^{\frac{2}{3}}\bigl(\frac{\sqrt{3 \Lambda}}{4} \,  t\bigr)\,,
\label{sm}
\een
which enjoys the remarkable property that the jerk equals 1,
\beq
j= a^2(\displaystyle\frac{d a}{dt})^{-3}
\frac{d^3a}{dt^3}=1\,.
\label{jerk}
\eeq
See Ref. \cite{Dunajski:2008tg} for details and original references.

Here we shall leave the  precise form of $a(t)$  unspecified.
The coordinate $t$ is called \emph{cosmic time}. The spatial coordinate $\bx$ is usually said to be \emph{comoving}
since the world lines of the cosmic fluid  have constant  $\bx $.
Two events simultaneous with respect to constant time,
i.e. with $x_1^{\mu} =(0, \bx_1) $ and $x_2^{\mu} =(0, \bx_2)$
have a time-dependent proper separation $a(t)(\bx_1-\bx_2)$.

\subsection{The choice of null vector field}

The vector field
\ben
l^\mu \frac{\p}{\p x^\mu} = g(t)
\frac{1}{\sqrt{2}} \bigl (a \frac{\p }{\p t} - \frac{\p} {\p {z}} \bigr )
\label{nullvector}
\een
where $g(t)$ is a nonvanishing  arbitrary function
is past directed  and null  but is neither covariantly constant nor Killing, as it can be checked by a tedious calculation. The associated one-form is
\ben
l_\mu dx^\mu =  -a^2 g \frac{1}{\sqrt{2}}
\bigl( \frac{1}{a} dt + d {z} \bigr )\,.
\label{nullform}
\een

If $\dot x^\mu =\frac{d x^\mu}{d\lambda}$ then
\ben
\cL= - m (a^2g)^b \bigl (\frac{1}{\sqrt{2} }
( a(t)^{-1}\dot t + \ \dot {z}) \bigr)^b
(\dot t^2 - a^2 \dot \bx ^2)^{\half (1-b) }
\label{Lag}
\een
 is a possible  Bogoslovsky-Finsler-type Lagrangian for a particle  of mass $m$.
It  admits three commuting symmetries  generated by $\frac{\p}{\p \bx} $
and hence three conserved momenta $ \bp = \frac{\p \cL}{\p \dot \bx} $\,.

If $b=0$, then Eq. (\ref{Lag}) is the standard action for a freely moving particle
in a flat isotropic Friedmann-Lema\^{\i}tre universe.

\subsection{Hubble friction}\label{frictionsec}

A notable feature of the free motion of a massive particle
moving in a Friedmann-Lema\^{\i}tre universe is \emph{Hubble friction}. The conserved momenta are
\ben
\bp= m a^2\, \frac{d \bx}{d\tau} \,
\quad\text{where}\quad
d \tau = \sqrt{1-(a\frac{ d\bx}{dt})^2}\;dt \,.
\label{VII7}
\een
Here $d \tau$ is  the increment of  proper time along the world line of a particle.
The four-velocity of the particle with respect to the local inertial reference frame $\frac{\p}{\p t}, \frac{\p}{a(t)\p \bx}$
is
\ben
\bu= a(t) \frac{d\bx}{d \tau}
\quad\text{whence}\quad\bu =
\frac{\bp}{m a(t)} \,.
\label{VII8}
\een

One may also define a velocity $\bv$ measured in units of cosmic time $t$,
$
\bv = a(t) \frac{d\bx}{dt},
$
so that
\ben
d\tau=\sqrt{1-\bv^2}\, dt \,,\quad \bu = \frac{\bv}{\sqrt{1- \bv^2 }}
\aand
\bv = \frac{\bp}{m} \frac{1}{\sqrt {a^2 + \frac{\bp^2}{m^2}}} \;.
\een
Hence,
\beq\fbox{$\;\displaystyle\frac{d\bx}{dt}=\displaystyle\frac{\bp}{a\,\sqrt{m^2a^2 + \bp^2}}\;.
$}
\label{xta}
\eeq

Thus, in an expanding phase in which $a(t)$ increases with time,
both $\bv$ and $\bu$ decrease with time.
However, as a consequence of isotropy, their directions remain constant.
The fact that we have three conserved momenta and the  constraint
\ben
\bigl(\frac{dt}{d \tau}\bigr )^2 = 1+ a^2 \bigl(\frac{d \bx }{d \tau  }\bigr)^2
\een
implies  that the system of geodesics is completely integrable.
In fact,
\ben
d\bx =
\frac{\bp}{ma^2}\frac{dt}{\sqrt{1+(\frac{\bp}{ma})^2}}
\;\aand\;
d\tau= \frac{ d t}{\sqrt{1+(\frac{\bp}{a m})^2 }} \,.
\een

\subsection{Conformal flatness}
Before proceeding further we recall that the Friedmann-Lema\^{\i}tre metric
(\ref{FLmetric}) is conformally flat, as becomes
clear if we define   \emph{conformal time} $\eta$ by
\ben
\eta(t) = \int^t\!\frac{d \tilde{t}}{a(\tilde{t})} \label{eta}
\een
where the lower limit is left unspecified for the time being.
In terms of cosmic time, the Friedmann-Lema\^{\i}tre metric (\ref{FLmetric}) becomes
\ben
 g_{\mu\nu} dx^\mu dx^\nu =  a^2 \Big \{- d\eta^2 + d{z} + dx^i dx^i\Bigr \}
= a^2 \Big \{2 du dv  + dx^i dx ^i \Bigr \}
\een
where $a^2$ is regarded as a function of conformal time $\eta$, and we introduce the light-cone coordinates
\ben
u =  \frac{{z}  +  \eta}{\sqrt{2}},
\qquad
 v = \frac{{z} -\eta}{\sqrt{2}}\,.
\label{uvdef}
\een
From Eq. (\ref{nullform}), we learn that
$
l_\mu dx^\mu  = - a^2 g du
$
whence our \B-F-ized line element  is
\ben
ds = a^{1+b} g^b\,\big(-2dudv - dx^i dx^i \big)^{\half (1-b)}\,(du)^b \,.
\label{FFLlineelem}
\een
This Lagrangian would yield Bogoslovsky's original flat spacetime model provided we choose $g(t)$ such that
\ben
f= a^{1+b} g^b =1.
\een
The only freedom with this model would be to introduce an arbitrary factor
\ben
\cL_f= f(\eta) \big(-2\dot u\dot v - \dot x^i\dot x^i\big)^{\half (1-b)}(\dot u)^b
\,,
\label{VII16}
\een
which amounts to saying that the mass depends upon cosmic time.

This situation is the same as in the ordinary spatially
flat  Friedmann-Lema\^{\i}tre cosmology for which $b=0$. We can either say the Universe is
expanding, but our rulers, are constructed  from massive particles, all of whose
masses are constant in cosmic time $t$,  or that the Universe is time independent, but the rulers all change with the same time dependence.
In that case, the phenomenon of Hubble friction would be ascribed not to the expansion of
the Universe but to masses are getting heavier.

\subsection{Redshifting}
If we adopt (\ref{VII16}) then
light rays move along straight lines in $(\eta, \bx)$  coordinates.
Emitters  and observers (e.g. galaxies and astronomers)  are usually held to be at rest in these coordinates.

Suppose the observer is at the origin at $(\eta_0, 0,0 ,0)$ and  receives  light rays from a galaxy
at $(\eta_e, x_e, y_e, z_e)$  so that the duration of emission  in conformal time is  $d \eta _e$
and the duration of the corresponding observation is $d\eta _e $, then
\ben
 d \eta_e =  d \eta_0\,.
\een
Then the emitted and observed proper times are
$
d\tau_e =f(\eta_e) d\eta_e\,,\, d \tau_0=f(\eta_0) d\eta_0
$
and so the redshift is
\ben
1+z = d\tau_0 /d\tau_e  = f(\eta_0)/f(\eta_e)\,.
\een
Thus, if the universe is expanding --- that is, if $f^\prime >0$ --- then the signal received is redshifted, and contrariwise if the Universe is contracting --- that is, if   $f^\prime <0$.

Note that under these assumptions, the emitted light from all galaxies at the same conformal time will be redshifted in the same way. That is, \emph{the redshift should be isotropic}.

\subsection{A possible choice for $f(\eta)$}\label{fetasec}
As mentioned earlier, our observed Universe is well
described by a scale factor $a(t)$
given by Eq. (\ref{sm}). Applying  the Einstein  equations
to the Friedmann-Lema\^{\i}tre  metric [Eq. (\ref{FLmetric})],
one finds that  it is supported  by a pressure-free fluid (some of it visible and some of it  not -- so-called dark matter) and a positive cosmological constant term $\Lambda$ often called dark energy.
Near the big bang   --- i.e., for small $t$ ---
$a(t)  \propto t^{2/3}$ , because the $\Lambda$ term is negligible.
This  is the Einstein-de Sitter model.
At late times, $a(t) \propto \exp{\sqrt{\Lambda/3}\, t}$,
which exhibits  cosmic acceleration. This is de Sitter spacetime.

From Eq. (\ref{eta}), choosing Eq. (\ref{sm})  and setting $ag=1$ in
Eq. (\ref{FFLlineelem}),  we have
\besub
\begin{align}
\eta(t) &=\int _0^t \sinh^{-\frac{2}{3}}
\bigl (\frac{\sqrt{3\Lambda}}{4}\tilde t \bigr) d \tilde t
\label{etat}
\\[6pt]
f(\eta) &= a(t) = \sinh^{\frac{2}{3}}\bigl(\frac{\sqrt{3 \Lambda}}{4} t  \bigr)\, ,
\label{f(eta)}
\end{align}
\label{fetaeta}
\esub
\begin{figure}[h]
\includegraphics[scale=.24]{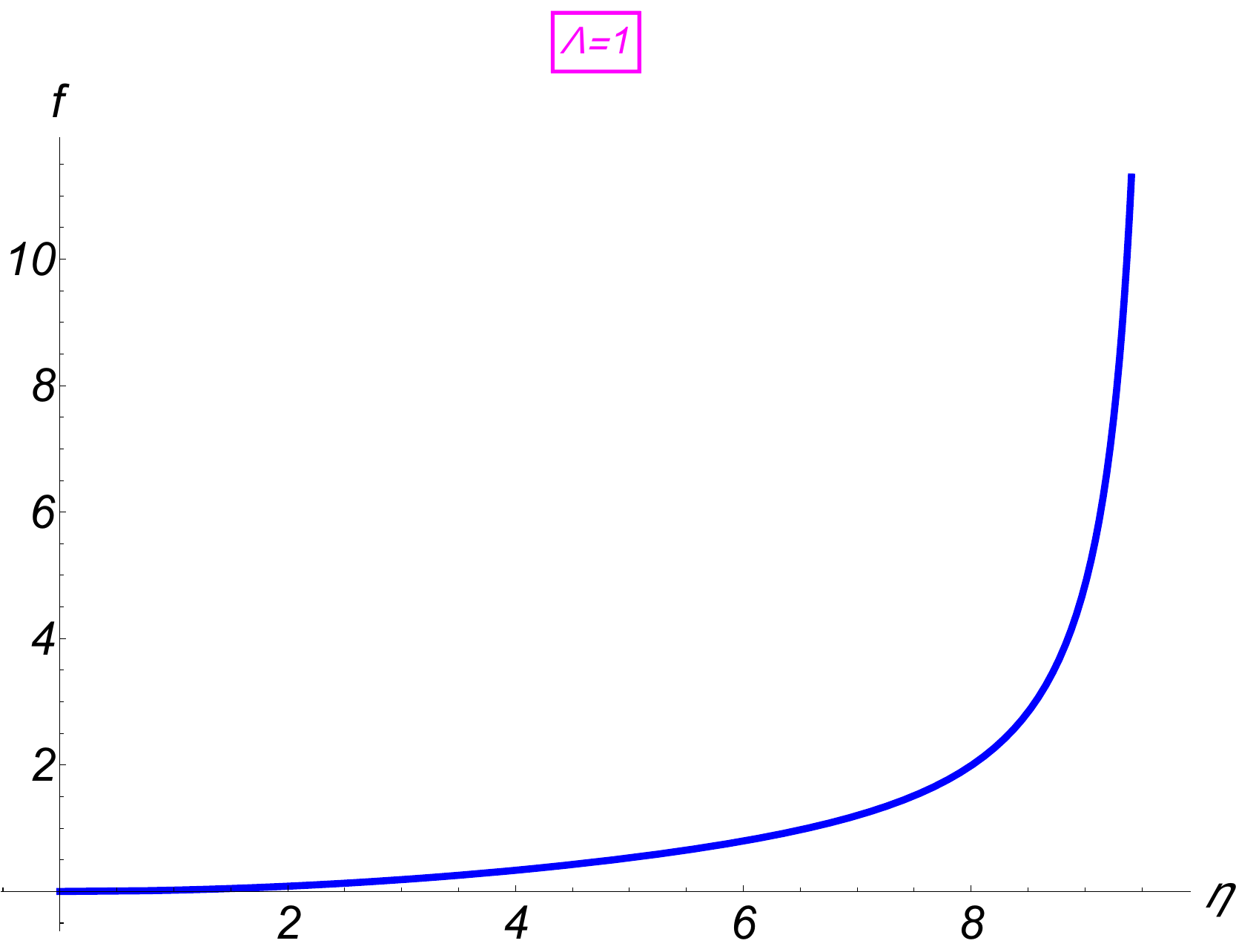}
\vskip-3mm
\caption{\textit{\small
The conformal factor \eqref{f(eta)} of the \FL  model \eqref{FLmetric}, expressed as function of the conformal time, $\eta$, obtained by numerical integration of \eqref{fetaeta}.
\label{fetaPIC}
}}
\end{figure}
This step depends only on the scale factor $a$ in \eqref{FLmetric} and does not involve the deformation parameter $b$. See Fig.\ref{fetaPIC}.
It is worth noting that conformal time as a function of cosmic time is bounded from above --- as  happens for de Sitter space, to which our spacetime tends when $t\to\infty$.

\section{Bogoslovsky-Finsler-Friedmann-Lema\^{\i}tre  Geodesics}\label{geosec}
Written in coordinates $(\eta,x,y,z)$, the Lagrangian
\eqref{VII16} is
\ben
\cL_f= - m  f(\eta) (\frac{\dot \eta + \dot  z}{\sqrt{2} })^b
( \dot  \eta ^2   - \dot  x ^2 -  \dot y ^2 - \dot z ^2 ) ^{\half (1-b)}\,,
\een
providing us with the momenta
\besub
\begin{align}
p_x &= m (1-b)f(\eta) \dot x \big(\frac{\dot \eta + \dot  z}{\sqrt{2} }\big)^b
\big(\dot\eta ^2  - \dot x^2 -  \dot y ^2 - \dot z ^2\big)^{-\half (1+b)}
\label{px}
\\[8pt]
p_y &= m (1-b) f(\eta) \dot y \big(\frac{\dot \eta + \dot  z}{\sqrt{2}}\big)^b
\big(\dot \eta^2 - \dot x^2 - \dot y^2 -\dot z^2\big) ^{-\half (1+b)}
\label{py}
\\[8pt]
p_z &= m (1-b)f(\eta)\,\dot z\,
\big(\frac{\dot\eta+\dot z}{\sqrt{2}}\big)^{b}
\big(\dot \eta^2-\dot x^2-\dot y^2-\dot z^2\big)^{-\half (1+b)}\nn\\
&\quad -\frac{b}{\sqrt{2}} mf(\eta)
\big(\frac{\dot\eta+\dot z}{\sqrt{2}}\big)^{-1+b}
\big(\dot\eta^2-\dot x^2-\dot y^2-\dot z^2\big)^{\half (1-b)}
\label{pz}
\\[8pt]
p_\eta  &=- m f(\eta)\big(\frac{\dot \eta + \dot z}{\sqrt{2}}\big)^{b-1}
\big(\dot \eta^2 - \dot x^2 - \dot y^2 - \dot z^2\big)^{-\half(1+b)}\nn
\\[3pt]
&\qquad\Bigl((1-b) \dot \eta (\frac{\dot\eta + \dot z}{\sqrt{2}}) +\frac{b}{\sqrt{2}}
\big(\dot\eta^2 - \dot x^2 + \dot y^2 - \dot z^2\big)\Bigr)
\label{peta}
\end{align}
\label{BFFLp}
\esub
Evidently the three momenta $p_x,p_y,p_z$ are conserved.
Moreover, since
\ben
\frac{p_y}{p_x} = \frac{dy}{dx}
\een
the projections of the geodesics onto the transverse $x-y$ plane are straight lines.
Choosing the proper time as parameter, $\lambda =\tau$, one has the constraint
\ben
  f(\eta) (\frac{\dot  \eta + \dot  z}{\sqrt{2} })^b
(\dot \eta ^2 - \dot x ^2 - \dot y ^2 - \dot z ^2 ) ^{\half (1-b)} =1
  \label{constraint}
 \een
which we may rewrite in terms of conformal time, $\eta$, as an equation for $\tau$ as
\ben
\tau ^\prime = f(\eta) (\frac{1+ z^\prime}{\sqrt{2} })^b ( 1   -  (x^\prime) ^2 -
( y^\prime) ^2 - (z^\prime)  ^2 ) ^{\half (1-b)}
  \label{propertime} \,.
\een
where $(x^\prime,y^\prime, z^\prime) = (\frac{dx}{d \eta}, \frac{dy}{d\eta},\frac{dz}{d\eta})$.

If $f^\prime =0$, then $p_\mu/f(\eta)$ is independent of $\eta$, leaving us with the same straight-line motion at constant velocity as for the flat Bogoslovsky spacetime.

As we have seen above, even if  $f^\prime \ne 0$, the projections
of the motion on the $x-y$ plane are straight lines,
although not with constant speed with respect to the conformal time
$\eta$. The speeds of the  projections
onto the $x-z$ and $y-z$ planes are also not at constant $\eta$ speed,
but  they are not straight lines either. Over
conformal $\eta$ times that are short compared
with $\frac{f}{f^\prime}$ they are approximately straight lines with slopes given by $\frac{p_x}{mf(\eta)}$
but over longer time periods, the speeds and directions change reflecting
precisely the effects of Hubble friction.

\begin{figure}[h]
\includegraphics[scale=.26]{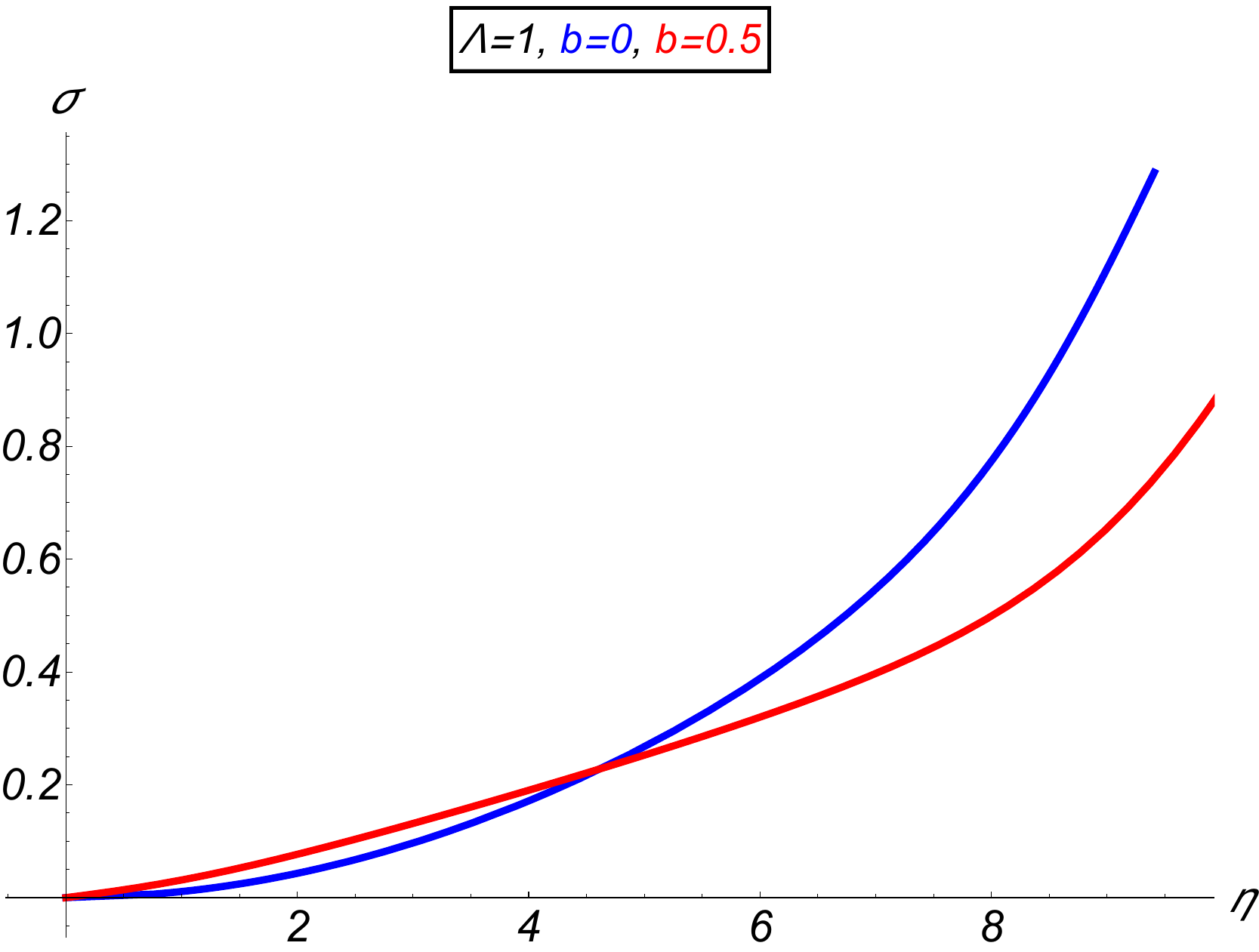}
\hskip10mm
\includegraphics[scale=.27]{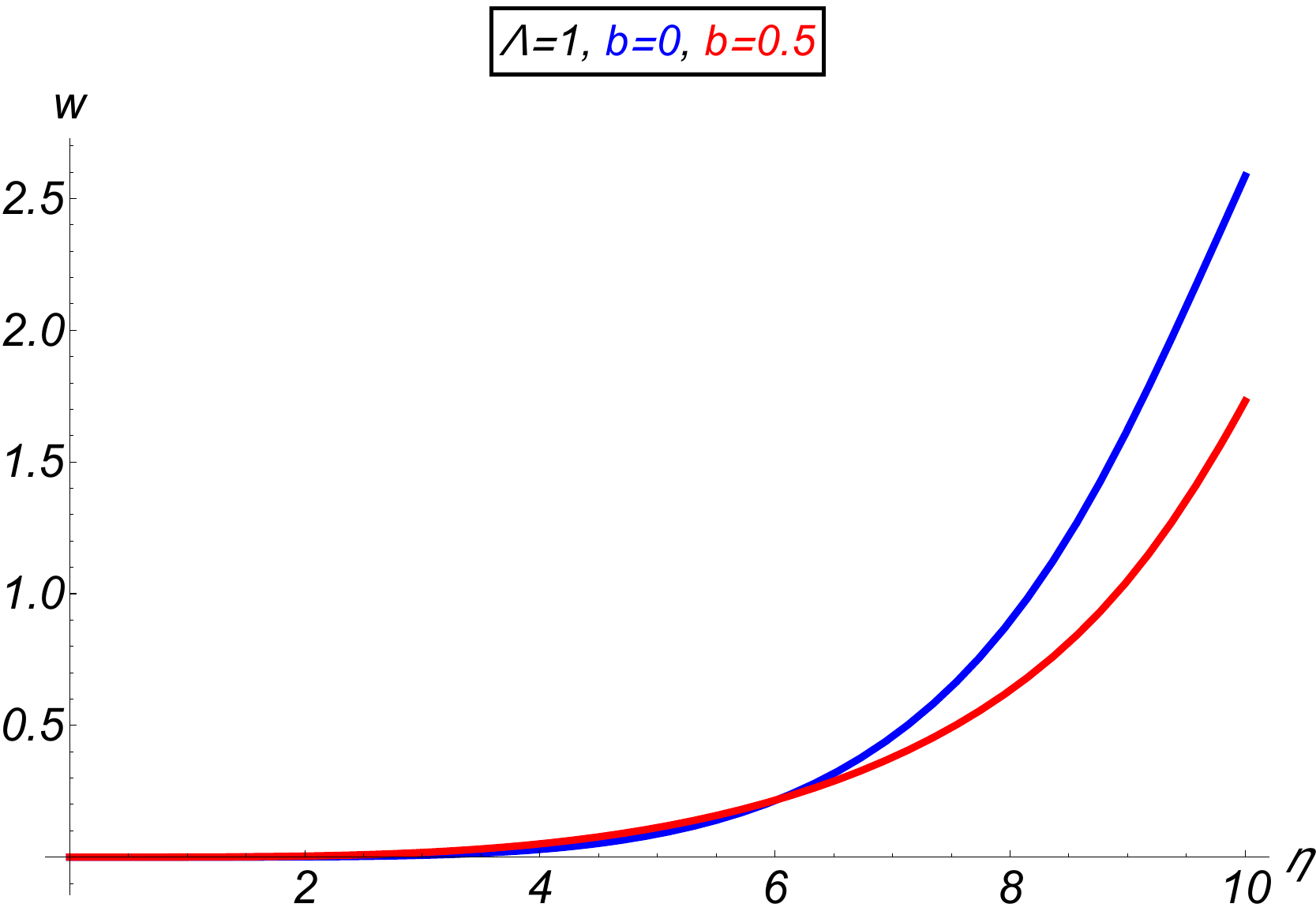}\\
\null\vskip-7mm\hskip-8mm
\null(a)\hskip70mm (b)\\
\vskip-3mm
\caption{\textit{\small (a) $\sigma(\eta)$ and (b) $w(\eta)$ in \eqref{wrep} plotted  for \textcolor{blue}{$\bf b=0$} and for  \textcolor{red}{$\bf b=0.5$}.
\label{sigmawb05}
}}
\end{figure}

The geodesics are conveniently studied by switching to conformal time, $\eta$.
Introducing
\begin{equation}
\label{uwrel}
 \dot{u} = \frac{\dot{\eta} + \dot{z}}{\sqrt{2}}
  \aand \dot{w} = \dot{\eta}^2 -\dot{x}^2 - \dot{y}^2 - \dot{z}^2
\end{equation}
in place of $\dot{\eta}$ and $\dot{z}$, Eqs. \eqref{px}, \eqref{py}, and \eqref{pz} become\vspace{-2mm}
\begin{subequations}
  \begin{align}
    m(b-1)f(\eta) \dot{x} \dot{u}^b+ p_x \dot{w}^{\frac{b+1}{2}} =0
    \\
    m(b-1) f(\eta) \dot{y} \dot{u}^b+ p_y \dot{w}^{\frac{b+1}{2}} =0 \\
    \label{eq3diff}
    \frac{m f(\eta)}{2\sqrt{2}} \dot{u}^{b-1} \dot{w}^{\frac{1}{2} (-b-1)} \left(2 (b-1) \dot{u}^2+(b+1) \dot{w}-(b-1) \left(\dot{x}^2+\dot{y}^2\right)\right) + p_z =0 .
  \end{align}
\end{subequations}
The first two equations imply identical evolution:
\begin{equation}
\label{sydot}
\dot{x} = \frac{\dot{u}^{-b} \dot{w}^{\frac{b+1}{2}}}{m(1-b) f(\eta)}\,p_x\,,
\qquad
\dot{y} = \frac{\dot{u}^{-b} \dot{w}^{\frac{b+1}{2}}}{m(1-b) f(\eta)}\,p_y
\end{equation}
which confirms once again that the transverse projection is a straight line, owing to $\dot{x}/\dot{y}=p_x/p_y=\const$
With their help, Eq. \eqref{eq3diff} becomes
\begin{equation}
\label{eq3diff2}
  \begin{split}
    -\sqrt{2} (b-1) m^2 f(\eta)^2 \dot{u}^{2 b} \left(2 (b-1) \dot{u}^2+(b+1) \dot{w}\right)-4 (b-1) m p_z f(\eta) \dot{u}^{b+1} \dot{w}^{\frac{b+1}{2}} \\
    +\sqrt{2} \left(p_x^2+p_y^2\right) \dot{w}^{b+1} =0 .
  \end{split}
\end{equation}
By reparametrizing $w(\lambda)$ as
\begin{equation}
\label{wrep}
  w(\lambda) = \int \!\! \sigma(\lambda)^2 \dot{u}(\lambda)^2 d\lambda,
\end{equation}
where $\sigma(\lambda)$ is a new function that we introduce, Eq. \eqref{eq3diff2} reduces from a differential one to an algebraic
\begin{equation}
\label{eq3alg}
  \begin{split}
    -\sqrt{2} (b-1) m^2f(\eta)^2 \left((b+1) \sigma(\lambda)^2 + 2 (b-1)\right) -4 (b-1) m f(\eta) p_z \sigma(\lambda)^{b+1} \\
    +\sqrt{2} \left(p_x^2+p_y^2\right) \sigma(\lambda)^{2 (b+1)} =0.
  \end{split}
\end{equation}
For $b=0$ this is simply quadratic in $\sigma(\lambda)$, but for $b\neq 0$ it is not trivial to solve it for $\sigma$, cf. Fig.\ref{sigmawb05}(a). However, as it is quadratic in $f(\eta)$, the inverse problem [which amounts to choosing $\sigma(\lambda)$  to find the corresponding $f(\lambda)$] still works. The  functions $\sigma(\eta)$ and $w(\eta)$ are plotted in Fig.\ref{sigmawb05}.
%
Using Eqs. \eqref{uwrel}, \eqref{wrep} and \eqref{sydot}  we get
\begin{subequations}
\label{foODEs}
\begin{align}
\label{foODE1}
  \dot{\eta} & = \frac{\dot{u}}{2 \sqrt{2}} \left(2 + \sigma^2 +\frac{\left(p_x^2+p_y^2\right)\sigma^{2(b+1)}}{(b-1)^2 m^2 f(\eta)^2}\right)
\\[7pt]
  \dot{x} & = \dot{u}\frac{\sigma^{b+1}}{(1-b)m f(\eta)}\,p_x
  \label{dxpx}
\\[7pt]
  \dot{y} & = \dot{u}\frac{\sigma^{b+1}}{(1-b)m f(\eta)}\,p_y
  \label{dypy}
    \\[7pt]
\label{foODE4}
  \dot{z} & = \frac{\dot{u}}{2 \sqrt{2}} \left(2-\sigma^2 -\frac{\left(p_x^2+p_y^2\right)\sigma^{2 (b+1)}}{(b-1)^2 m^2 f(\eta)^2} \right)
\end{align}
\end{subequations}
together with the algebraic constraint between $\sigma(\lambda)$ and $f(\eta(\lambda))$ in Eq. \eqref{eq3alg}.

The joint system can be shown to satisfy the Euler-Lagrange equations.

The $u(\lambda)$ that remains unspecified in \eqref{foODEs} and disappears from Eq. \eqref{eq3alg} serves as a gauge parameter [by seeing the ratios of derivatives that are being formed in Eq. \eqref{foODEs}], for which we can simply set $u(\lambda)=\lambda$. So in this time-gauge $\eta(\lambda) + z(\lambda) = \sqrt{2} \lambda$, which is compatible with Eqs. \eqref{foODE1} and \eqref{foODE4}, as seen above.
We thus have, in the ``conformal-time gauge,''
\begin{subequations}
\begin{align}
  \frac{d x}{d\eta} & = \frac{2 \sqrt{2} (1-b) m \sigma^{b+1} f(\eta)}{(1-b)^2 m^2 \left(\sigma^2+2\right) f(\eta)^2+\left(p_x^2+p_y^2\right) \sigma^{2 (b+1)}}\, p_x
\\[8pt]
  \frac{d y}{d\eta} & = \frac{2 \sqrt{2} (1-b) m \sigma^{b+1} f(\eta)}{(1-b)^2 m^2 \left(\sigma^2+2\right) f(\eta)^2+\left(p_x^2+p_y^2\right) \sigma^{2( b+1)}}\, p_y
\\[8pt]
  \frac{dz}{d\eta} & = \displaystyle\frac{2-\sigma^2 -\displaystyle\frac{\left(p_x^2+p_y^2\right)\sigma^{2 (b+1)}}{(1-b)^2 m^2 f(\eta)^2}}{2+\sigma^2+\displaystyle\frac{\left(p_x^2+p_y^2\right)\sigma^{2 (b+1)}}{(1-b)^2 m^2 f(\eta)^2}}
 \,.
\end{align}
\label{xyzeta}
\end{subequations}
%
\kikezd {\textbf{The \FL case  $b=0$}}

If $b=0$, the algebraic relation \eqref{eq3alg} is quadratic and can be solved for $\sigma$:
\beqa
  \sigma(\eta) = \pm
\frac{\sqrt{2}\, m f(\eta)}{\sqrt{m^2 f(\eta)^2+\bp^2}\pm p_z}\,,
\label{b0sigma}
\eeqa
shown by the blue line in  Fig. \ref{sigmawb05}(a) for the upper sign \footnote{Choosing the lower sign would amount to an overall sign change when that of $p_z$ is also reversed.}.
In terms of conformal time $\eta$
\begin{equation}
  \fbox{$ \displaystyle\frac{d\bx}{d\eta} =
 \displaystyle{\pm}\frac{\bp}{\sqrt{m^2f(\eta)^2+\bp^2}}
  \;. $}
\label{b0xeta}
\end{equation}
to be compared  with Eq. \eqref{xta}.
The consistency with the equations in Sec.\ref{frictionsec} follows from
\ben
d\tau= \frac{mf^2d\eta}{\sqrt{m^2f^2+{\bp}^2}},
\quad
 d\bx = \frac{d\eta}{\sqrt{m^2f^2 + {\bp}^2}}\,\bp
 \Rarrow
\fbox{$ \displaystyle\frac{d\bx}{d\tau}=\displaystyle\frac{1}{mf^2}\,\bp\;. $}
\label{b0eqn}
\een

\begin{figure}[h]
\includegraphics[scale=.25]{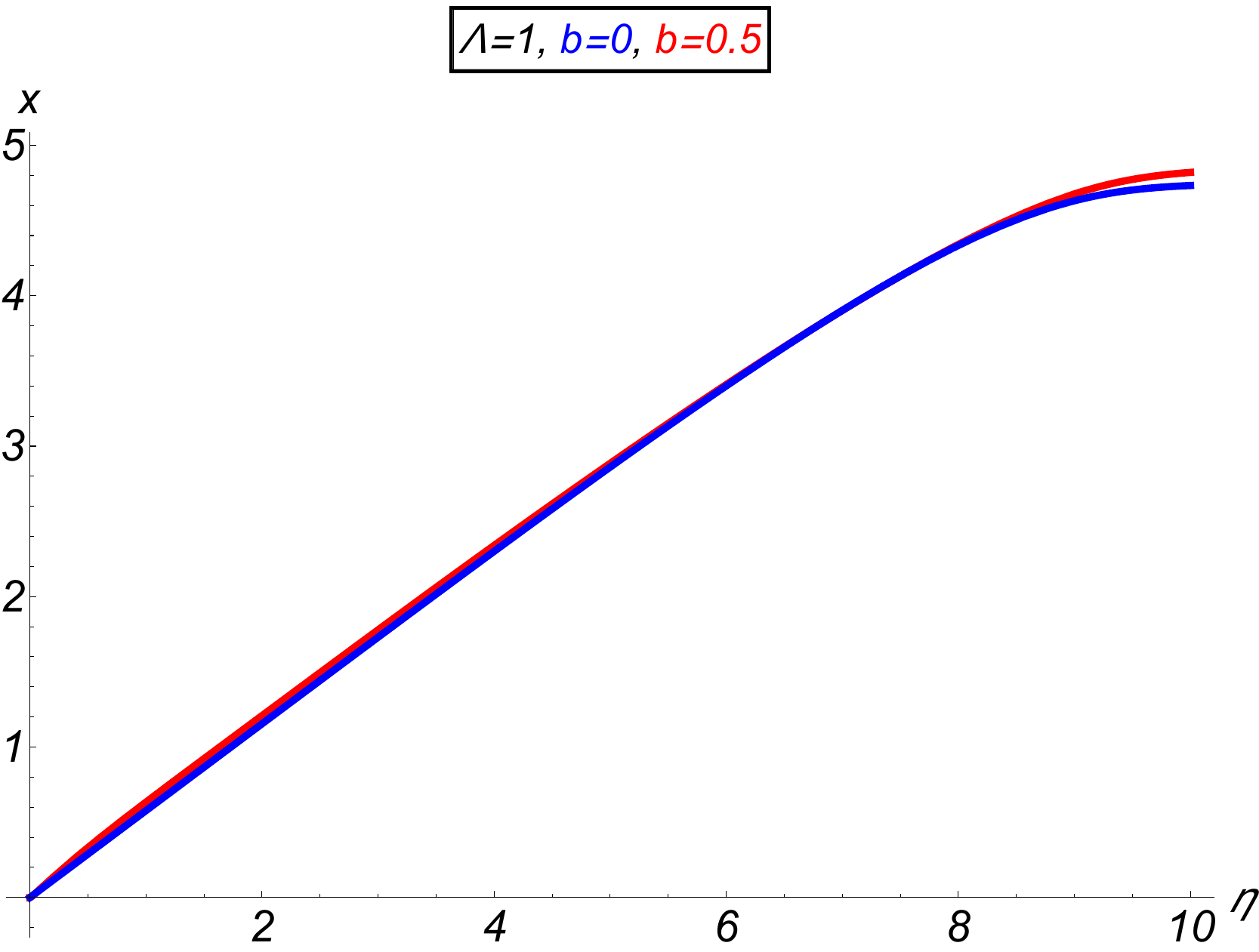}
\vskip-3mm
\caption{\textit{\small For \textcolor{blue}{${\bf b=0}$} all trajectories follow straight lines and have identical evolution. For \textcolor{red}{${\bf b=0.5}$} $z(\eta)$ become different  from the transverse trajectories ($x(\eta), y(\eta)$)    consistently with \eqref{xyzeta}, as shown in Fig.\ref{xzplaneb5}.
\label{xtraj05}
}}
\end{figure}

We could not obtain analytical expressions; however, using the
  numerically calculated values of $f(\eta)$ (see Fig.\ref{fetaPIC}) allows us to plot $x(\eta)$ by solving Eq. \eqref{b0xeta}, as shown in Fig.\ref{xtraj05} \footnote{The two signs in \eqref{b0sigma} can be compensated by $\bp \to-\bp$ implying an overall sign change. In what follows the upper sign will be chosen.}.

\begin{figure}[h]
\includegraphics[scale=.22]{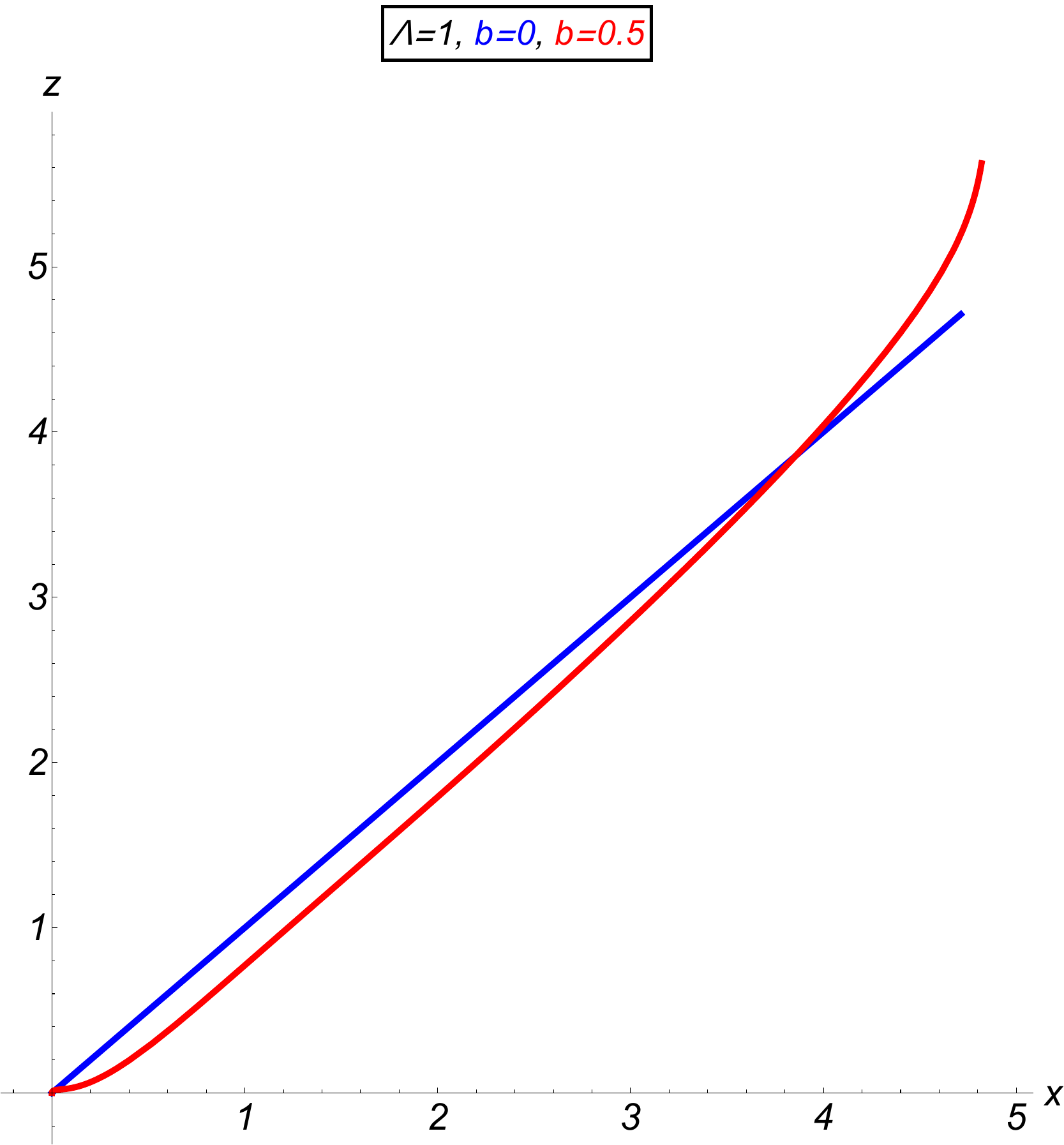}
\hskip11mm
\includegraphics[scale=.29]{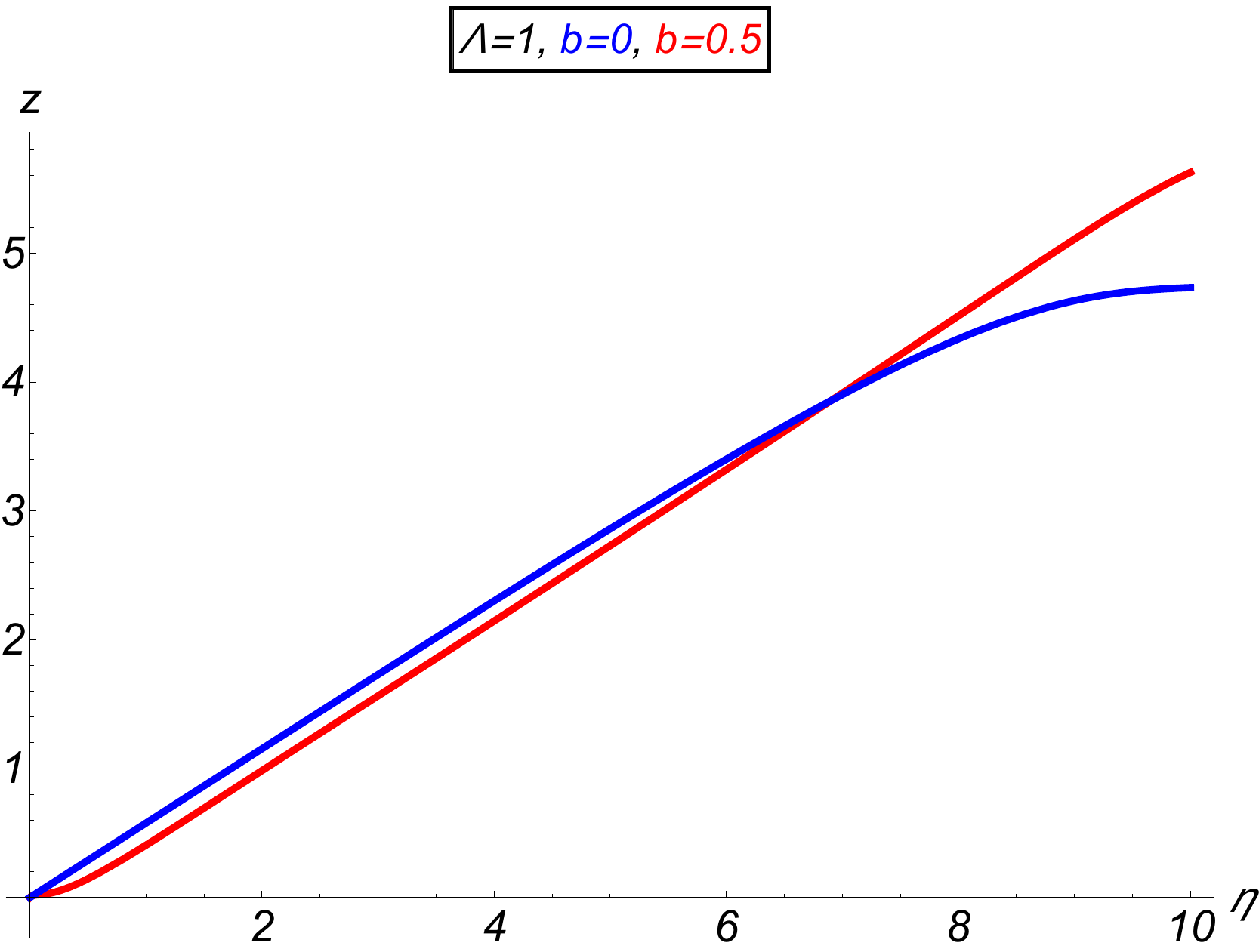}\\
\vspace{-8mm}
\null\hskip-9mm(a)\hskip70mm (b)
\vskip-4mm
\caption{\textit{\small For {$\textcolor{red}{{\bf b}>0}$} the motion in the $x-z$ plane is not more along a straight line (as it is for \textcolor{blue}{${\bf b}=0$}). The Hubble friction slows down the $z$ motion for \textcolor{blue}{${\bf b}=0$} but not when \textcolor{red}{${\bf b}>0$}.
\label{xzplaneb5}
}}
\end{figure}
\begin{figure}[h]
\includegraphics[scale=.3]{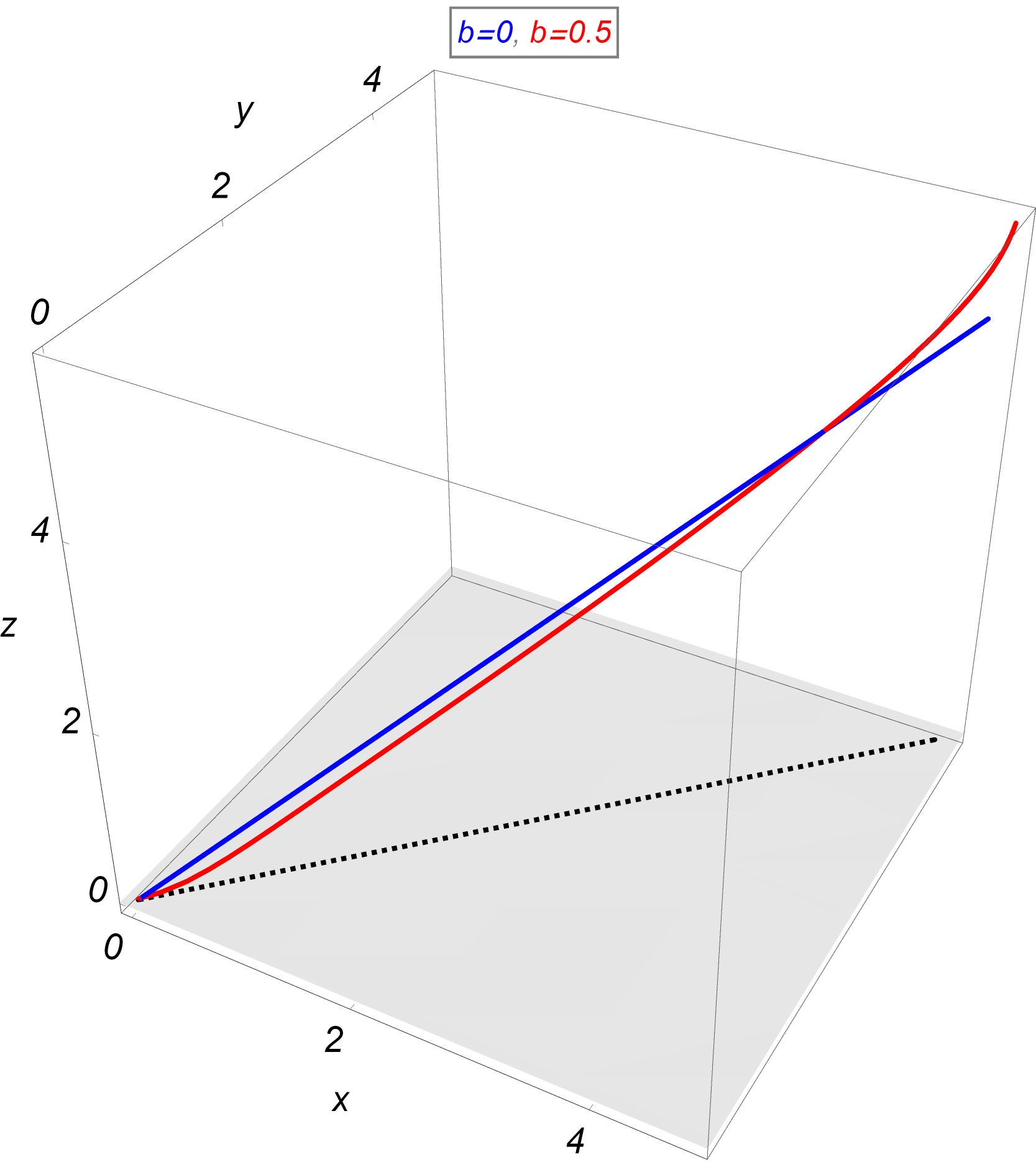}
\vskip-3mm
\caption{\textit{\small
For the {\FL model} for {${\bf b=0}$} the $3D$
trajectory is a straight line. For  the {\B-F} modification {${\bf b=0.5}$} however, while the projection to the $x-y$ plane is still along a straight line, the $z$-component becomes curved, consistently with Fig.\ref{xzplaneb5}.
\label{xyztrajb5FL}
}}
\end{figure}

The $b=0$ case nicely illustrates
\emph{Hubble friction} : all trajectories slow down and ultimately come to rest. For $b \ne 0$  it seems that this happens only for the transverse motion but not for the motion in the $z$ direction, see Fig.\ref{xyztrajb5FL}.
The slowing down in the transverse case is plausible from Eqs. \eqref{dxpx} and \eqref{dypy}.

\section{Conclusion}

In this paper, motivated by work by  Bogoslovsky \cite{B1,B2,B3,B4}, and by that of Tavakol and Van den Bergh, and Roxburgh \cite{Tavakol0,Tavakol,Roxburgh}, and by
  Cohen and Glashow \cite{Glashow}, and more recently by others \cite{Kostelecky11,Fuster1,Fuster2,Minguzzi,Perlick,Edwards,Bernal,Fuster20}, we have studied the free motion of a massive particle moving in a one-parameter family of Finslerian deformations
of a plane gravitational wave. By free motion, we mean  that it
extremizes the proper time along its timelike world line.
Finslerian proper time is measured by replacing the usual square-root integrand
$$
\sqrt{-g_{\mu\nu}\frac{dx^\mu}{d\lambda} \frac{dx^\nu}{d\lambda}}
\;\qquad\text{with}\qquad
\big(-g_{\mu\nu}l^{\mu}\frac{dx^\nu}{d\lambda}\big)^b
\big(-g_{\mu\nu}\frac{dx^\mu}{d\lambda}
\frac{dx^\nu}{d\lambda}\big)^{\half (1-b)}\,,
$$
where $l^\mu$ is a null vector field and  $b$ is a dimensionless constant.

In earlier work, we have shown that because of the  five-dimensional isometry group of plane gravitational waves,  the motion of   the usual timelike geodesics is completely integrable.

In the present paper, we have shown
that the five-dimensional partially broken Carroll  symmetry group $G_5$ remains a symmetry of our  Finslerian line element  provided we choose  the null vector field $l^\mu$ to be the covariantly constant null vector of the underlying  gravitational  wave.
As a consequence, we find that that not only is the free motion completely integrable, but it differs only in that the  ``vertical''
coordinate $v$ involves in turn a $b$-dependent term, which is linear in the retarded time coordinate $u$.
The motion in the transverse directions is unchanged. The situation is analogous to what happens for massive vs massless geodesics in a pp-wave \cite{nonlocal}.


The symmetry of the  \B-F model is in fact that of the very special relativity (VRS) type ;
in the Minkowski case it is the eight-parameter $\DISIM_b(2)$   \cite{MaxSIM,B3,Silagadze}. The clue is to deform a U-V boost $N_0$ to $N_b$  as in Eq.  \eqref{Nb}.
The trick works for certain nontrivial profiles, as for the $U^{-2}$ discussed at the end of Sec.\ref{Carrollsec}.

We have also examined the free motion of a   Finslerian
deformation of a homogeneous pp-wave which is an  Einstein-Maxwell solution.
The resulting spacetime is a Cahen-Wallach symmetric space \cite{Cahen}
and arises in a wide variety of physical applications and whose null reduction in the fashion of Eisenhart  and Duval et al. \cite{Bargmann} is a simple harmonic oscillator with a Newton-Hooke-type symmetry. Here again, the free motion
is qualitatively  independent of the deformation parameter $b$.

We have also studied a simple anisotropic cosmological model
based on that of  Friedmann and Lema\^{\i}tre  with vanishing spatial curvature.
Because  the latter is conformally flat, the motion of massive particles is equivalent to motion in  flat Bogoslovsky spacetime except that all masses become time dependent with identical time dependence.

Although our present Universe shows little sign of anisotropy
of the sort that arises in Bogoslovsky-Finsler metrics, that may not have been true earlier in the history of the Universe since the absence of anisotropy now is usually
ascribed to a rapid phase of inflation during  which the scale  factor of the Universe increased by a factor of perhaps 60 $e$-folds. It is of interest, therefore to study geodesics in   Bogoslovsky-Finsler deformations of \FL metrics.

\begin{acknowledgments}
We would like to thank Ettore Minguzzi for correspondence.
This work was partially supported by the National Key Research and Development Program of China (No. 2016YFE0130800) and the National Natural Science Foundation of China (Grant No. 11975320).
\end{acknowledgments}


\end{document}